\documentclass[fleqn,usenatbib]{mnras}

\usepackage[T1]{fontenc}
\usepackage{ae,aecompl}

\usepackage{graphicx}	% Including figure files
\usepackage{amsmath}	% Advanced maths commands
\usepackage{amssymb}	% Extra maths symbols
\usepackage{caption}
\usepackage{pdflscape}
\usepackage{subcaption}
\usepackage[normalem]{ulem}
\usepackage{mathptmx}

\newcommand{\cz}{\mathrm{c}}
\newcommand{\sz}{\mathrm{s}}

\usepackage{bm}
\usepackage{wasysym}
\usepackage{upgreek}

\title[High resolution ALMA and HST imaging of $\kappa$\,CrB]{High resolution ALMA and HST imaging of $\kappa\,$CrB: a broad debris disc around a post-main sequence star with low-mass companions}

\author[J. B. Lovell et al.]{J. B. Lovell$^{1}$\thanks{E-mail: jl638@cam.ac.uk / joshualovellastro@gmail.com}, M. C. Wyatt$^{1}$, P. Kalas$^{2,3}$, G. M. Kennedy$^{4,5}$, S. Marino$^{1}$, A. Bonsor$^{1}$, \newauthor Z. Penoyre$^{1}$, B. J. Fulton$^{6,7}$ N. Pawellek$^{8,9}$ \\
$^1$Institute of Astronomy, University of Cambridge, Madingley Road, Cambridge, CB3 0HA, UK\\
$^2$University of California, Berkeley, 501 Campbell Hall, 3411 Berkeley, CA 94720-3411 \\
$^3$Institute of Astrophysics, FORTH, GR-71110 Heraklion, Greece \\
$^4$Department of Physics, University of Warwick, Coventry, CV4 7AL, UK \\
$^5$Centre for Exoplanets and Habitability, University of Warwick, Gibbet Hill Road, Coventry CV4 7AL, UK\\
$^6$California Institute of Technology, Pasadena, CA 91125, USA \\
$^7$IPAC-NASA Exoplanet Science Institute Pasadena, CA 91125, USA \\
$^{8}$Department of Astrophysics, University of Vienna, T\"urkenschanzstra\ss{}e 17, 1180, Vienna, Austria \\
$^{9}$Konkoly Observatory, Research Centre for Astronomy and Earth Sciences, Konkoly Thege Mikl\'os \'ut 15-17, H-1121 Budapest, Hungary}

\date{Accepted XXX. Received YYY; in original form ZZZ}

\pubyear{2022}

\usepackage{newtxtext,newtxmath}

\begin{document}
\label{firstpage}
\pagerange{\pageref{firstpage}--\pageref{lastpage}}
\maketitle

\begin{abstract}
$\kappa\,$CrB is a ${\sim}2.5$\,Gyr old K1 sub-giant star, with an eccentric exo-Jupiter at ${\sim}2.8$\,au and a debris disc at tens of au.
We present ALMA Band~6 (1.3\,mm) and HST scattered light (0.6$\,\mu$m) images, demonstrating $\kappa\,$CrB's broad debris disc, covering an extent $50{-}180\,$au in the millimetre (peaking at $110$\,au), and $51{-}280\,$au in scattered light (peaking at 73\,au). 
By modelling the millimetre emission, we estimate the dust mass as ${\sim}0.016\,M_\oplus$, and constrain lower-limit planetesimal sizes as $D_{\rm{max}}{\gtrsim}1\,$km and the planetesimal belt mass as $M_{\rm{disc}}{\gtrsim}1\,M_\oplus$.
We constrain the properties of an outer body causing a linear trend in 17\,years of radial velocity data to have a semi-major axis $8{-}66$\,au and a mass $0.4{-}120\,M_{\rm{Jup}}$. 
There is a large inner cavity seen in the millimetre emission, which we show is consistent with carving by such an outer massive companion with a string of lower mass planets. 
Our scattered light modelling shows that the dust must have a high anisotropic scattering factor ($g{\sim}0.8{-}0.9$) but an inclination ($i{\sim}30{-}40^\circ$) that is inferred to be significantly lower than the $i{\sim}61\degr$ millimetre inclination.
The origin of such a discrepancy is unclear, but could be caused by a misalignment in the micron and millimetre sized dust.
We place an upper limit on the CO gas mass of $M_{\rm{CO}}{<}(4.2{-}13) \times10^{-7}\,M_\oplus$, and show this to be consistent with levels expected from planetesimal collisions, or from CO-ice sublimation as $\kappa\,$CrB begins its giant branch ascent.

%We constrain the upper limit on the CO gas mass between $4.2{-}13\times10^{-7}\,M_\oplus$, and show this to be consistent with production from either CO-rich planetesimal collisions and/or CO-ice sublimation. 
\end{abstract}


%%%%%%%%%%%%%%%%%%%%%%%%%%%%%%%%%%%%%%%%%%%%%%%%%%
%%%%%%%%%%%%%%%%% BODY OF PAPER %%%%%%%%%%%%%%%%%%

\section{Introduction}
\label{sec:intro}
The debris discs that have been observed around other stars are inferred to be dust, produced in the collisional cascades of exo-asteroid and exo-comet belts.
Impacts between large belt objects form smaller planetesimals and observable quantities of debris dust at infrared and sub-millimetre/millimetre wavelengths \citep{Wyatt08, Hughes18}.
Debris discs have been observed with features such as cavities, gaps, warps, clumps, radial asymmetries, and gas (such as Carbon Monoxide, CO) as a result of belt collisions and/or interactions between planets and belts \citep{Kral17, Marino18, Matra19, Faramaz19, Lovell21c}.
Consequently, such features mean that studies of debris discs can constrain broader aspects of planetary systems such as their architecture and evolution that may not otherwise be observable.

Exo-Jupiter planets, which we define here as a class of giant planet with $0.1{-}5\,M_{\rm{Jup}}$ and orbital radii between 1-5\,au, are observed around ${\sim}$5\% of stars \citep{Winn15}.
Such planets are regularly determined to have eccentric orbits \citep{Chiang13}, an outcome predicted to emerge from planetary system instabilities \citep{Juric08} and/or the Kozai mechanism \citep{Nagasawa08}.
Debris discs at tens of au can be depleted by eccentric exo-Jupiters and simultaneously influence other closer-in planets dynamically \citep{Gomes05, Raymond11, Raymond12}, yet the presence of exo-Jupiters appears thus far to have little to no impact on the occurrence of detectable debris discs \citep{Bryden09, MoroMartin15, Yelverton20}. 
Despite this, exo-Jupiters and other outer massive planets can strongly influence the structure of planetesimal belts over long Myr-Gyr timescales \citep[see e.g.,][]{Mustill09} that can be imaged with high resolution observatories, such as ALMA and HST.
Recently \citet{Lovell21c} presented new ALMA images of the debris disc around q$^1$~Eri, a F9 main sequence star that hosts an exo-Jupiter, and connected the disc morphology to a plausible planetary system architecture.
%discussed the plausibility of the asymmetric disc morphology as the result of carving by a planet following outward migration (for example, following scattering via the inner giant planet).
%Such outward planet migration could be common in systems with exo-Jupiters,
Studying the morphology of debris discs at high resolution around stars hosting exo-Jupiters can therefore offer avenues to explore the planetary architecture and exoplanetary system evolutionary histories.

\begin{figure}
    \includegraphics[width=0.5\textwidth]{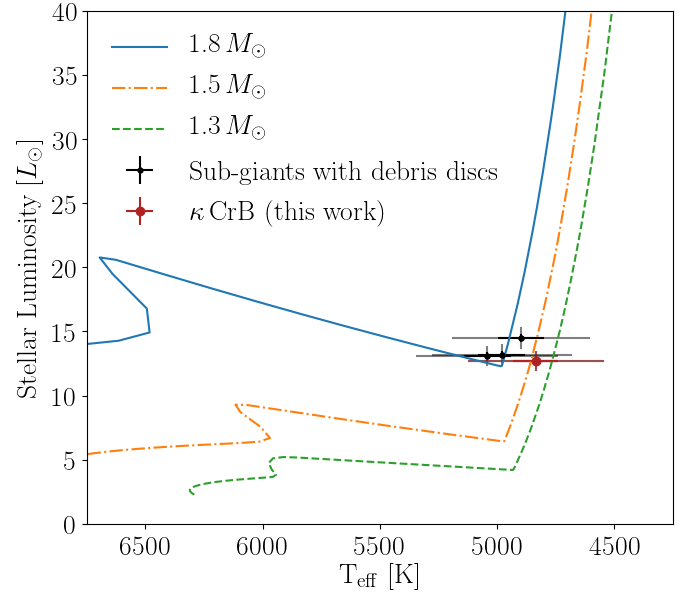}
    \caption{HR diagram showing three stellar tracks \citep[from the models of][]{Hurley00} for stars with $Z{=}0.025$ and masses of 1.8, 1.5 and 1.3$M_\odot$, respectively for the solid-blue, dash-dotted-amber and dashed-green lines and the four known post-main sequence sub-giants that host debris discs \citep{Bonsor14}. Error bars represent 1$\sigma$ and 3$\sigma$ uncertainties (assuming $\sigma{=}2\%$, see $\S$\ref{sec:SED}). All four are located near the base of the giant branch based on their ``Priam'' and ``FLAME'' modelled effective temperatures and luminosities \citep[i.e., from][]{Gaia18}. All four are consistent with A-type stellar evolution tracks.}
    \label{fig:HRdiag}
\end{figure}

Debris discs are brighter and have higher detection fractions around main sequence A-type stars, spanning their main sequence lifetimes \citep[with fractional occurrence rates of ${\sim}20\%$ over the range 5--850\,Myr, see e.g.,][]{Rieke05, Su06, Wyatt07}.
Therefore the planetary systems of retired A-type stars provide useful laboratories to investigate long-term (${\gtrsim}\,$Gyr) planet-disc interactions, for the old evolved systems that host at least one known planet and at least one observable debris disc.
As part of a study of 36 such evolved post main sequence stars, \citet{Bonsor14} presented evidence of four sub-giants hosting a far-infrared excess indicative of a debris disc (observed at 100\,$\mu$m and 160\,$\mu$m with Herschel), of which three host at least one confirmed giant planet (HR\,8461, HD\,131496, and $\kappa\,$CrB, all of which were once A-type stars).
At a distance of 30.1\,pc \citep{Gaia18}, of the known sub-giants that host at least one planet and debris disc, $\kappa\,$CrB is the closest (by a factor $2{-}4\times$), and hosts the brightest disc.
Although all four known sub-giants with debris discs are at the base of the giant branch (see Fig.~\ref{fig:HRdiag}), $\kappa\,$CrB's closer distance and far-infrared disc brightness makes it the most ideal system to investigate the long-term effects of planet-belt interactions around a post-main sequence star. 

$\kappa\,$CrB (HD 142091, HIP 77655) is a K1 sub-giant star (an evolved A-type main sequence star), with a mass ${\sim}1.8\,M_\odot$ and age ${>}2.5$ Gyr \citep{Johnson08}.
$\kappa\,$CrB has been known to be an evolved sub-giant for many decades \citep[based on spectral photometry, surface gravity, and absorption features, see e.g.,][]{Keenan80, McWilliam90}.
It hosts an exo-Jupiter, $\kappa\,$CrB\,b, with semimajor axis $a_{\rm{pl}}=2.8\,$au, mass $M_{\rm{pl}}\,\sin{i}=1.6\,M_{\rm{Jup}}$ and eccentricity $e_{\rm{pl}}=0.07$ \citep[see][]{Johnson08}. 
$\kappa\,$CrB's debris disc was first imaged at 100 and 160 $\mu$m at resolutions of $6.7''$ and $11.4''$ respectively with \textit{Herschel} PACS observations \citep{Bonsor13}.
The data were fitted with a disc with modelled inclinations between $59{-}60\degr$ and position angles between $142{-}145\degr$, and either a wide (20--220\,au) dust belt or two belts centered at 41 and 165\,au.
Although broad morphological constraints were placed on the disc, the low resolution imaging meant that a number of questions were left open; is this a single or multiple component debris disc; and how have possible disc-planet interactions carved the disc? 

We present here the first ALMA and HST images of $\kappa\,$CrB, providing strict constraints on this planetary system, with observational resolution ${\gtrsim}3\times$ higher than previous Herschel images in an attempt to unravel the evolution of this planetary system. 
In $\S$\ref{sec:dataRed} we detail our methodology for reducing our ALMA and HST data sets, in $\S$\ref{sec:dataAnalysis} we present our observational analysis of this system, in $\S$\ref{sec:modelling} our methodology to model the data, in $\S$\ref{sec:COflux} derive upper limits on the CO flux, and in $\S$\ref{sec:planetArch} we derive limits on the putative planetary architecture. 
In $\S$\ref{sec:discussion} we provide a detailed discussion of our results and their possible interpretations, and summarise our key findings and conclusions in $\S$\ref{sec:conclusions}.

\begin{figure*}
    \includegraphics[width=1.0\textwidth]{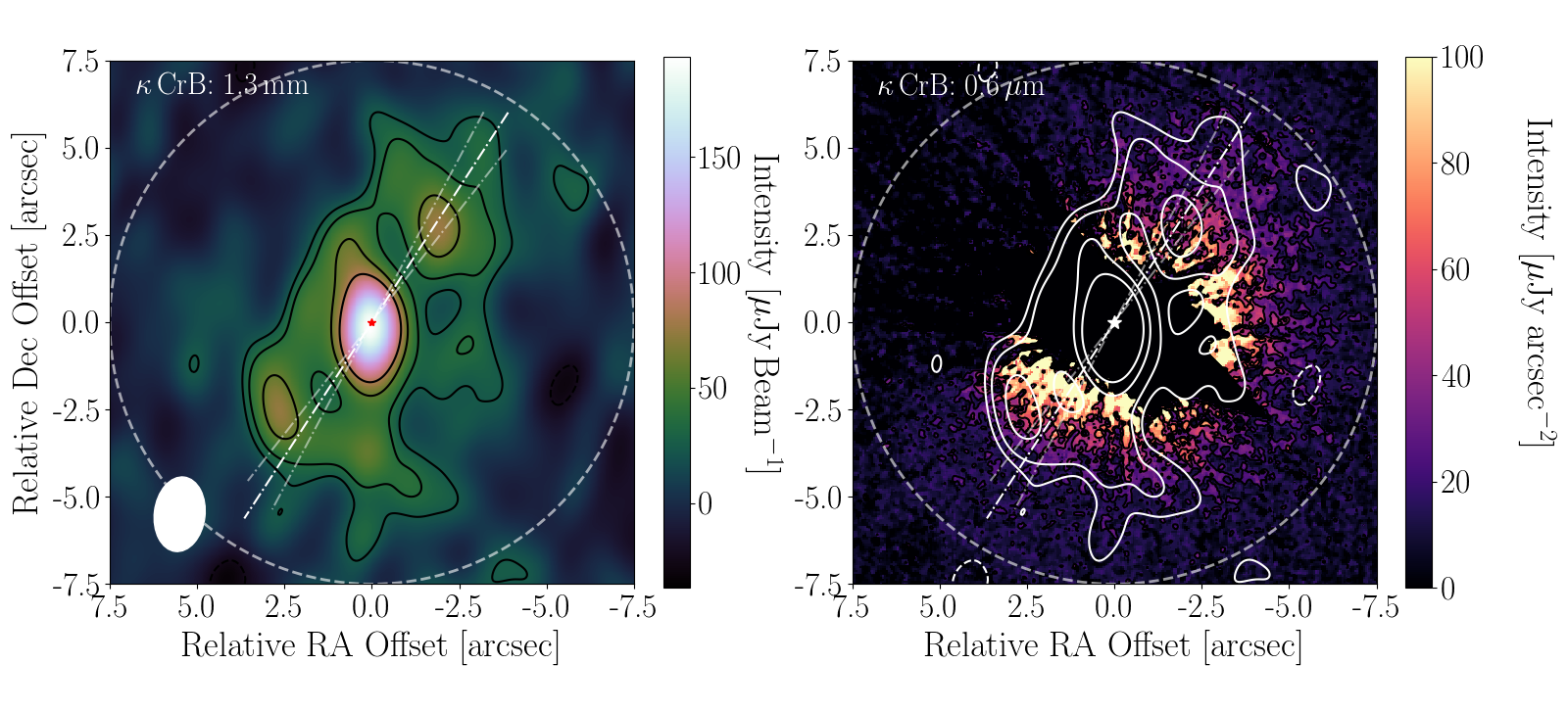}
    \caption{High resolution images of the $\kappa$\,CrB system. In both, North points up, East points left, a 7.5'' dashed ring is plotted, dash-dot lines indicate the position angles ($146{\pm}5^\circ$) and stellar locations are marked in the image centres with asterisk symbols (red left, white right). Left: Band~6 (1.3\,mm) natural weight cleaned image, showing the star, the disc, and outer emission. This image has an rms of ${\sim}12.0\,\mu$Jy\,beam$^{-1}$, and contour lines of ${\pm}$2, ${\pm}$3, ${\pm}$5 and ${\pm}$7$\,\sigma$ emission. Right: HST scattered light image at optical wavelengths showing dust nebulosity coincident with the ALMA emission, with an image rms of ${\sim}8\,\mu$Jy\,arcsec$^{-2}$, with ALMA Band~6 contours overlaid. }
    \label{fig:kCrB_B6_HST}
\end{figure*}

\section{Data Reduction}
\label{sec:dataRed}
\subsection{ALMA data reduction}
\label{sec:datareductionALMA}
$\kappa$\,CrB was observed for ${\sim}195$ minutes on source in four scheduling blocks with ALMA in Band~6 (1.1--1.4 mm) during Cycle~7, using 49 antennas with minimum and maximum baselines ranging from 15.0 to 312.7\,m (project 2019.1.01443.T, PI: Lovell). 
Each scheduling block observed for ${\sim}$49 minutes on source, and were conducted on the 4-Dec 2019, 20-Dec 2019, and 2x 21-Dec 2019. % (i.e., over the complete set of ALMA observations, the system remained effectively stationary on the sky).
In all cases, the correlator had 3 spectral windows centred on frequencies of $228.527$, $215.527$, and $217.527$\,GHz, for continuum observations with a bandwidth of $2.000$\,GHz and channel widths of $15.625$\,MHz, as well as a spectral window centred on a topocentric frequency of $230.569$\,GHz with a bandwidth of $1.875$\,GHz and channel widths of $976.562$\,kHz for CO J=2-1 spectral line observations. 
The average frequency of these observations corresponds to our quoted wavelength of 1.34\,mm (which herein we quote as 1.3\,mm).
The visibility data set was calibrated using the \textit{CASA} software version 5.6.1-8 with the standard pipeline provided by the ALMA Observatory. 
The $\it{plotms}$ task in $\it{CASA}$ was used to examine the visibility amplitudes as functions of both time and uv-distance, and as a result it was deemed that no data flagging would be necessary. 
The measurement set was purged of calibrator data and time averaged into 10-second bins, channel averaged into 4 channels per spectral window, and the \textit{CASA} $\it{statwt}$ task was run to estimate the visibility weights based on the measured variance.
Continuum imaging was conducted using the \textit{CASA} $\it{tclean}$ algorithm with natural weighting (providing a good trade-off between resolution and signal-to-noise), shown in Figure~\ref{fig:kCrB_B6_HST} (left) in which emission from both the central star and the inclined debris disc are clearly detected. 
The synthesised beam size in this image is $2.1 \times 1.4''$ (with a beam position angle $\rm{PA}_{\rm{Beam}}{=}173.1^{\circ}$, anti-clockwise from North), i.e., a physical resolution of $63.2\times 42.1$\,au which has clearly resolved the disc over multiple beams.

\subsection{HST data reduction}
\label{sec:datareductionHST}
$\kappa$\,CrB was observed with HST/STIS in three orbits on UT 2014-02-03. With the 50CORON aperture this 1024$\times$1024 CCD camera delivers unfiltered visible-light images with 0.05077 $''$\,pixel$^{-1}$.
The star was occulted by the WEDGEA0.6 and WEDGEA2.0 coronagraphic position which have widths 0.6$''$ and 2.0$''$, respectively.
The former provides close-in imaging but saturation limits the total integration time.
In each orbit we acquired 12 images with 2.5\,sec integration each at WEDGEA0.6.
For deeper, wide-field imaging we acquired nine images with 130\,sec integrations each at WEDGE2.0 during each orbit.
To facilitate subtraction of the optical point spread function (PSF), the telescope position angle was changed by 17$\degr$.00 between the first orbit and the second orbit, and 11$\degr$.83 between the second and third orbit.
A PSF reference star, HD~139323 (K3V), was observed with STIS at the WEDGE0.6 and WEDGE2.0 positions in a fourth orbit.

We registered the geometrically corrected, cosmic ray rejected files (*sx2.fits) by subtracting all images for a given wedge position from the very first image obtained in each orbit.
By iteratively shifting the images by 0.01\,pixel we determined the x,y shift that minimized the PSF residuals after subtraction.
To subtract the sky background, we determined the median pixel value in a 100$\times$100\,pixel box located in the top left corner of each image.
Every image was then divided by the integration time to provide units of counts\,sec$^{-1}$\,pixel$^{-1}$.
The images at each wedge position in each orbit were then median combined.

Each final $\kappa$\,CrB image for each wedge position in each orbit was then subtracted by the final image for each wedge position in the orbit that targeted HD~139323.
The x,y position and flux scaling of HD~139323 relative to $\kappa$\,CrB was iteratively adjusted to minimize the subtraction residuals in the annular region containing the PSF halo and telescope diffraction spikes.
Additionally, the WEDGEA2.0 data for $\kappa$\,CrB have an electronic or scattered-light artifact appearing as a positive-valued horizontal band across the entire detector to the left and right of the occulted star.
The band encompasses $\sim$37 rows and it is uniformly brighter (median value 0.07 cts\,sec$^{-1}$\,pix$^{-1}$) to the left of the star than to the right (0.04 cts\,sec$^{-1}$\,pix$^{-1}$) as measured in the PSF-subtracted images.
The steps to convert between cts\,sec$^{-1}$\,pix$^{-1}$ and $\mu$Jy\,arcsec$^{-2}$ are provided by \citet{Schneider16}, for which we note that with a gain of 4 and the pixel size stated above yields 1\,cts\,sec$^{-1}$\,pix$^{-1}{=}177\,\mu$Jy\,arcsec$^{-2}$. 
We subtracted the horizontal band using these values for rows 662:702 and 666:700 to the left and to the right of the star, respectively.
Before PSF subtraction, the intersections of the diffraction spikes were used to determine the stellar locations behind the occulting wedge positions.
The stellar location was then used as a center of rotation for the PSF-subtracted images to be oriented with north up and east left.
These north-rotated images were then median combined to provide two final PSF-subtracted images of $\kappa$\,CrB at the WEDGEA0.6 and WEDGEA2.0 positions representing 90\,seconds and 3510\,seconds cumulative integration time, respectively.

%\begin{figure*}
%    \includegraphics[width=1.0\textwidth]{Figures/radProf_r_az_together_230322.png}
%    \caption{Left: radial profile (in red) along major axis for the Band~6 data, along with curves for the stellar and disc peaks (blue and amber respectively) convolved with the beam, and a black dash-dot line representing $5\,\sigma$ emission. Right: azimuthal profiles around the disc, within three annuli (with radial extents of 60\,au, each relative the dotted lines above which these are plotted). In both plots the shaded regions represent $1\,\sigma$ error. }
%    \label{fig:kCrB_radProf}
%\end{figure*}
\begin{figure}
    \includegraphics[width=1.0\columnwidth]{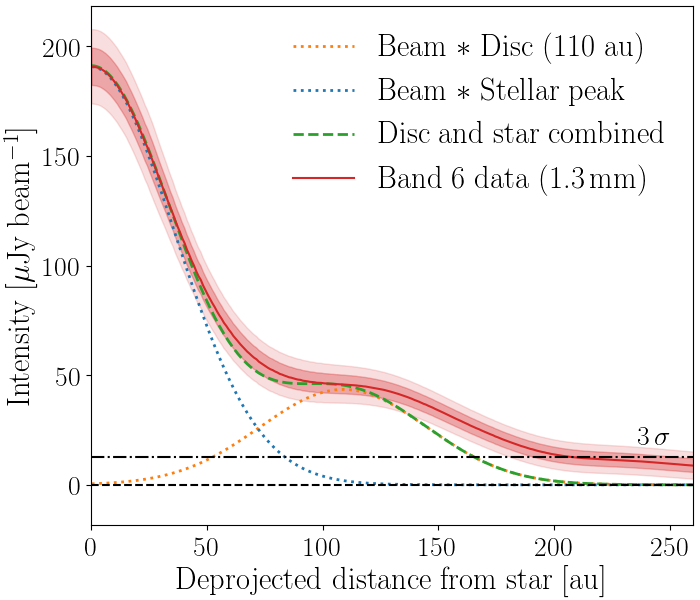}
    \caption{Azimuthally-averaged radial profile for the Band~6 data (red, solid), along with curves showing emission coincident with the star and the disc at a deprojected distance of 110\,au convolved with the beam (blue and amber respectively, dotted), their combination (green, dashed), and a black dash-dot line representing $3\,\sigma$ emission. The shaded regions represent $1\,\sigma$ and $2\,\sigma$ errors. }
    \label{fig:kCrB_radProf}
\end{figure}

Neither STIS image of $\kappa$\,CrB shows evidence for distinct morphological features such as rings or belts surrounding the star.
The WEDGEA0.6 data have negative values up to 1.7$''$ radius, and then positive values out to 4.1$''$ radius.
The negative halo may be a processing artifact: changes in the thermal environment of the telescope between orbits changes the PSF such that the PSF reference star is mismatched with that of the science target.
The morphology of the positive halo is elliptical with a major axis in the southeast-northwest direction.
However, because the WEDGEA0.6 position is near the edge of the detector, the $\kappa$\,CrB image is vignetted at 2.7$''$ radius in the PA range $45\degr{-}90\degr$.

Our analysis will therefore focus on the WEDGEA2.0 data which are not vignetted, and are deeper, but have a larger inner working angle (IWA; $\geq$1.7$''$ or 51\,au) than the WEDGEA0.6 data.
This image (Fig.~\ref{fig:kCrB_B6_HST}, right) does not have a negative-valued inner halo.
Instead, there is a positive halo with a flux deficit northeast of the star, producing an overall fan-like morphology.
A key concern is whether or not the deficit could be due to the occulting bar of the camera that crosses the field from the northeast to the southwest.
This is an unlikely explanation because the bar would also produce a light deficit symmetrically across the star between the northeast and southwest sides.
Also, the positive halo cannot be due to a residual of mismatched PSF's because such residuals would be azimuthally symmetric.
Debris discs inclined to the line of sight can exhibit fan-like morphologies when surrounded symmetrically by dust, with an asymmetric scattering phase function \citep{kalas96}.
We thus conclude that the fan-like light morphology after PSF subtraction is due to light scattered by the dust grains surrounding $\kappa$\,CrB.

\section{Observational Analysis}
\label{sec:dataAnalysis}
%In this section we present in turn our analyses of $\kappa\,$CrB's ALMA and HST images, and the system's complete flux distribution to assess the debris disc morphology.

\subsection{ALMA analysis}
\label{sec:ALMAobsAnalysis}
All analysis presented in this section is derived from the image of $\kappa$\,CrB as shown in the left panel of Fig.~\ref{fig:kCrB_B6_HST}, and we start by producing profiles of the emission.
However, in order to produce accurate radial profiles, we first require estimates of the disc position angle ($\rm{PA}$) and inclination ($i$). 
We start by measuring the $\rm{PA}$ measured anti-clockwise from North centred on the star by plotting the total flux within $5^\circ$ wedges within a distance of 250\,au (to enclose all disc emission from the stellar centre as a function of angle between 0 and 180$^\circ$ in $2^\circ$ increments) and found this was maximised for a $\rm{PA}{=}146{\pm}5^\circ$, consistent with previous work \citep[i.e., $142^\circ-145^\circ$ as modeled by][]{Bonsor13}.
We define the line parallel with the position angle through the centre of the star as the disc major axis, plotted on both panels of Fig.~\ref{fig:kCrB_B6_HST} along with its associated uncertainty.
Given the inclined nature of this disc, the brightness of the emission coincident with the star and the beam size, emission along the minor axis is fainter, and not well resolved. % (deeper imaging in the future might better measure dust emission in these regions).
To measure the inclination, we therefore subtracted the unresolved stellar emission using a 2D Gaussian, defined by the beam major and minor axes, beam position angle, and peak pixel intensity, from the centre of the Band~6 image and plotted elliptical rings around the residual emission, for which $i{=}\arccos{\rm{(major\,extent / minor\,extent}}$). 
We estimate the disc inclination as $61{\pm}5^\circ$, by varying the inclination such that all ellipses remained within the $3\,\sigma$ contour lines \citep[also consistent with previous work, i.e. $\sim$59$^\circ$ as modeled by][]{Bonsor13}.
Whilst these inclination and position angle values come with imaging uncertainty, we deem these sufficiently accurate for the purposes of our observational analysis.
In $\S$\ref{sec:modelling} with more detailed modelling we show that these are consistent with best fits of the interferometric visibilities.

Fig.~\ref{fig:kCrB_radProf} shows the azimuthally-averaged radial profile centred on the star, using our derived inclination of $61^\circ$ and position angle of 146$^\circ$.
Clear from this profile is that the peak emission is centred on the star, and that disc emission extends from ${\sim}$50--205\,au (1.7$''$--6.8$''$) based on the shoulder feature which falls below the $3\sigma$ level at ${\sim}$205\,au, but peaks at around ${\sim}110\,$au (3.7$''$).
To place constraints on the stellar and disc emission, over-plotted are the profiles of i) a point source with the same brightness as the central peak located at the origin (as a blue dotted line), and ii) a narrow ring of emission with the same brightness as the average belt emission at a deprojected radius of 110\,au (as an amber dotted line), with both having been convolved with the average beam extent.
Since the star is unresolved, the emission peak at this location can be used to constrain the 1.3\,mm stellar flux as $191{\pm}8\,\mu$Jy, which combined with the 10\% flux calibration uncertainty of ALMA results in $F_{\rm{\star,\,B6}}{=}191{\pm}20\,\mu$Jy.

The combined disc and star profile in Fig~\ref{fig:kCrB_radProf} shows that whilst a simple point source located at the stellar location and a narrow ring at 110\,au can broadly account for the emission internal to ${\sim}150$\,au, beyond this distance the outer edge emission is too bright, i.e., a narrow ring cannot fully interpret the disc towards $\kappa$\,CrB.
This could be the case if the disc is broad. 
We discuss further the potential origins of this emission signature at the end of this section.
It can be seen in the left plot of Fig.~\ref{fig:kCrB_B6_HST} that there is a faint clump of emission at an approximate relative RA-Dec location of [-1.0,-5.0] (i.e., at an approximate deprojected distance of ${\sim}205\,$au), with a few tens of $\mu$Jy\,beam$^{-1}$ brightness.
Given that the radial profile is consistent with noise at all radii beyond ${\sim}$205\,au this clump appears too faint to have significantly increased our estimate of the disc's extent.
We discuss the possible origins of this emission in \S\ref{sec:discussionClump}.

\begin{figure}
    \includegraphics[width=0.475\textwidth]{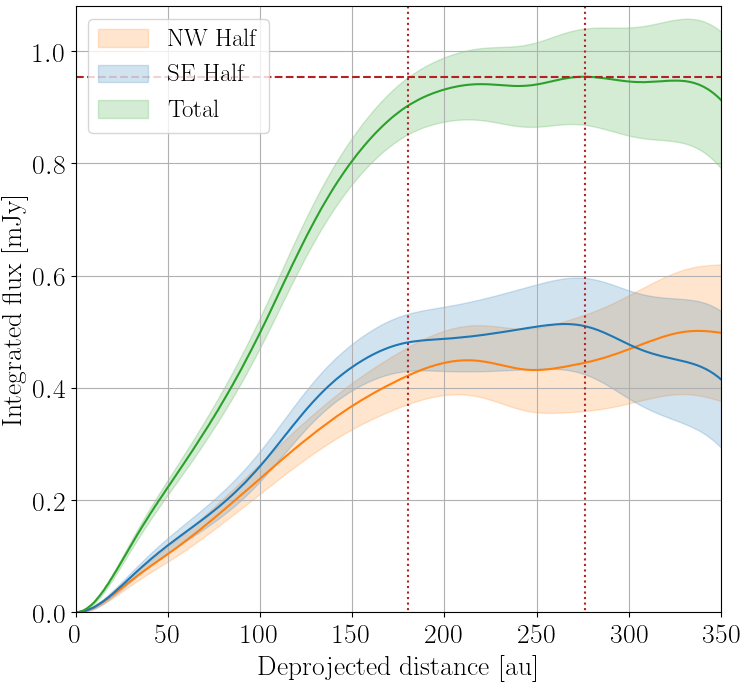}
    \caption{Band~6 flux profile as a function of deprojected radius of $\kappa\,$CrB. Highlighted regions represent $1\,\sigma$ error. 
    Here, the total flux can be measured. Split either side of the disc minor axis, this plot demonstrates that the total flux is symmetric about the star at the depth observed. }
    \label{fig:kCrB_fluxProf}
\end{figure}

The same conclusions from Fig.~\ref{fig:kCrB_radProf} can be drawn from the integrated flux profile shown in Fig.~\ref{fig:kCrB_fluxProf}. %, \citep[produced in a consistent manner as the analysis of q$^1$ Eri, see][]{Lovell21c}. 
Here we show that the integrated flux rises to a peak value at ${\sim}$276\,au, consistent within 1$\,\sigma$ with the integrated flux beyond 180\,au.
Although the average disc intensity falls below $3\sigma$ beyond ${\sim}$205\,au in Fig.~\ref{fig:kCrB_radProf}, given that the integrated flux profile is broadly constant beyond 180\,au, we interpret 180\,au as the disc outer edge in the millimetre.
Fig.~\ref{fig:kCrB_fluxProf} further demonstrates that at the depth of our observations there is no detectable asymmetry in the total flux in the two disc ansae, given that the integrated flux profiles in the NW and SE halves of the disc are consistent over all radii. 
The flux profile can additionally be used to measure the total 1.3\,mm flux from $\kappa$\,CrB based on the peak of the `Total' value of $0.96\pm0.04\,$mJy. 
Combined in quadrature with an assumed $10\%$ ALMA calibration uncertainty, this results in a value of $F_{\rm{tot,\,B6}}{=}0.96\pm0.10\,$mJy (which we include in Table~\ref{tab:SEDflux}, and later use to model the flux distribution in $\S$\ref{sec:SED}).

\subsubsection{Summary of ALMA analysis}
\label{sec:obsALMAsum}
In summary, we conclude from this ALMA analysis that having detected the debris disc towards $\kappa\,$CrB, we have observationally constrained its position angle on the sky as ${\rm{PA}}{=}146{\pm}5^\circ$, its inclination as $i{=}61{\pm}5^\circ$, the radial peak location of its disc emission at ${\sim}$110\,au, starting at ${\sim}50\,$au, and extending outwards to at least ${\sim}$180\,au. 

There are a number of ways that $\kappa\,$CrB's disc emission could arise based on the resolution of our imaging which we will explore by modelling the dust belt in $\S$\ref{sec:modelling}.
Whilst our observational analysis suggests that the broad disc emission arises from a belt of planetesimals, it remains unclear how broad the underlying planetesimal distribution may be.
We will test this with simple models that parameterise the surface density distribution of dust as a function of the radial distance from the star. 
These models will be able to 1) constrain the extent of the belt, and by comparing different parametrisations, 2) search for the presence of any underlying radial sub-structure in the disc.
We note that whilst broad debris discs can be modelled in many ways, here we will use just two parametrisations. 
We either model the disc with i) a radial power-law distribution (i.e., `model SPL', with free parameters for the inner and outer edges) or ii) a radial Gaussian distribution (i.e., `model SGR', with free parameters for the radial peak and width/standard deviation of the Gaussian).
In conjunction, these will offer sufficient insight to explore the debris disc and underlying planetesimal belt of $\kappa\,$CrB, which we fully described in $\S$\ref{sec:modelling}. 

\begin{figure}
    \includegraphics[width=0.475\textwidth]{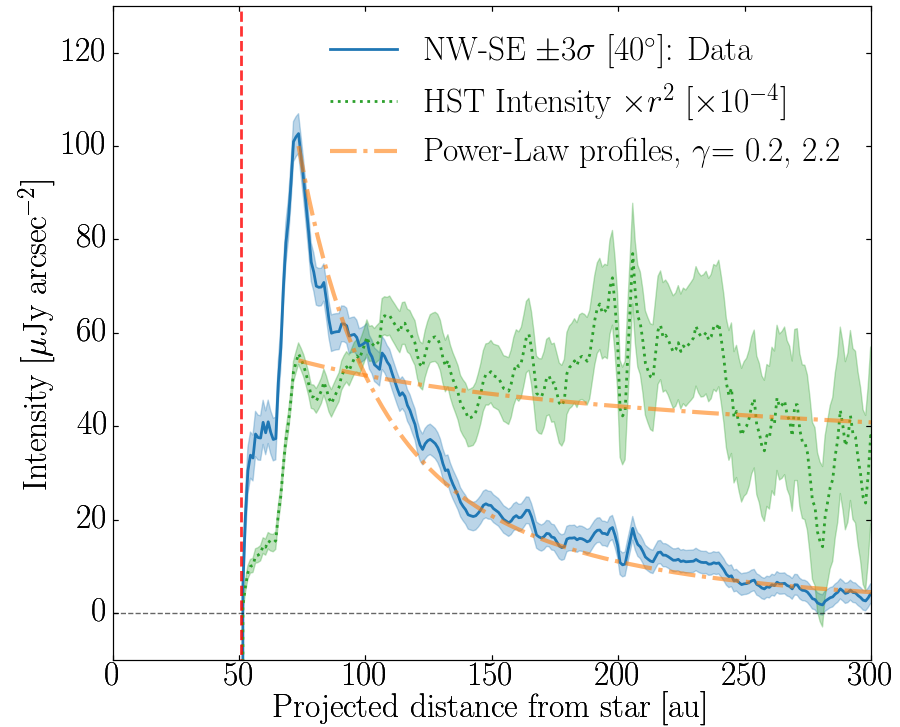}
    \caption{Radial profile for the HST scattered light emission (blue, solid), the same but scaled by $r^2$/10,000 (green, dotted), and modelled power-law profiles (amber, dashed), showing the decaying nature of the emission, over a broad radial distance from the star. The errors in scattered light data are ${\pm}3\,\sigma$, and the vertical red dashed line is at the radial location within which coronagraphic obscuration dominates the emission. }
    \label{fig:scatLightProfile}
\end{figure}

\subsection{HST analysis}
\label{sec:HSTobsAnalysis}
All analysis in this section is derived from the HST scattered light image of $\kappa\,$CrB as shown in the right panel of Fig.~\ref{fig:kCrB_B6_HST}.
We start by noting that all modelling of the scattered light data will be conducted in $\S$\ref{sec:modelHST}, and our comparison of the mm and scattered light observations will be discussed in $\S$\ref{sec:discussion}.
To first investigate the morphology of the disc that can be inferred from the image, we produced a radial profile of the scattered light by averaging emission within 10$^\circ$ wedges either side of the star, assuming a position angle of 146$^\circ$ and an inclination of 40$^\circ$ (which we discuss further and justify in $\S$\ref{sec:modelHST}).
This profile is shown in Fig.~\ref{fig:scatLightProfile}.
We note that the derived profiles with 10$^\circ$ wedge angles are consistent with a wide range of assumed inclinations and position angles, including those derived from the ALMA image. 
Such consistency demonstrates that there are large uncertainties present when estimating the geometry of discs from scattered light emission. 
The radial profile (designated `NW-SE Data') allows us to estimate the peak brightness, and the inner and outer-edges of the dust-scattered light. 
Together these can constrain the dust emission extent.
Further, since the surface brightness of scattered light images is proportional to the dust opacity, surface density of micron-sized grains, the scattering phase function and stellar irradiation, by scaling the surface brightness by $r^2$, the resulting profile corrects for the stellar irradiation \citep[which follows an inverse square law, see e.g.,][]{Stolker16}, and is thus a better approximation of the surface \textit{density} distribution of the micron-sized grains.
We therefore plot this same radial profile scaled by $r^2$.
Finally, we estimate these two emission profiles with decaying power-law models, which we discuss further below, and return to in $\S$\ref{sec:modelHST}.

Analysing Fig.~\ref{fig:scatLightProfile} in detail, we start by noting the WEDGEA2.0 IWA at 51\,au limits our ability to analyse emission within this radius (shown with a red dashed line). 
The brightness profile shows an initial `shoulder' of emission (between 51-67\,au), a sharp brightness peak (at 73\,au, with intensity ${\sim}100\,\mu$Jy\,arcsec$^{-2}$) and a decaying intensity profile outwards to ${\sim}280$\,au.
In terms of the $r^2$-corrected profile, the dust distribution has the same shoulder, and follows a relatively flat power-law profile between ${\sim}73-280$\,au, beyond which this becomes consistent with noise.
Based on the unknown true inner edge of the dust we conclude that the scattered light from circumstellar dust has a wide extent spanning ${\sim}51-280$\,au.

The radial profiles allow us to set two additional broad constraints on the dust distribution, and albedo which we discuss in turn here.
Firstly, starting at the 73\,au peak, by over-plotting an intensity profile $I(r) = I_0 \times (r/R_{\rm{in,\,HST}})^{-\gamma}$ (where $I_0$ is set to the inner-edge intensity, i.e., $I_0{=}100\,\mu$Jy\,arcsec$^{-2}$) and $I(r)\times r^2$, we show that the brightness and surface density fall-offs are consistent with simple power-laws.
We show these for a $\gamma{\sim}0.2$ for the surface density profile, and equivalently $\gamma{\sim}2.2$ for the surface brightness profile, from which we conclude that over the extent of the 73-280\,au emission, the fall-off in scattered light \textit{emission} is pronounced, with a dust surface density consistent with being relatively flat.
%This can be the case for example, if the dust particles are being continually produced in either a narrow or a broad belt and blown on to eccentric orbits (due to the radiation pressure of the star) with different distributions of eccentricity (i.e., since it is difficult to distinguish between the scattered light emission of a broad belt with low eccentricity dispersion and a narrow belt with high eccentricity dispersion).
Secondly, from the peak scattered light brightness we estimate the required albedo of the dust \citep[using equation 2 of][i.e., $S=fFA/ ((2 \pi \times \phi \times d\phi \times \cos{i}) (1-A))$, where S is the surface brightness in mJy\,arcsec$^{-2}$, $f$ is the fractional luminosity, $F$ is the stellar flux in mJy, $\phi$ is the disc semi-major axis in arcsec, $d\phi$ is the disc width in arcsec, $i$ is the inclination, and $A$ is the albedo]{Marshall18}. We find $A{\sim}10\%$, based on $F{=}2923$\,mJy, $f{=}5\times10^{-5}$ (see $\S$\ref{sec:intro}), $d\phi{=}$2-8\arcsec, and $i{=}40^\circ$ (we note $A$ varies by just ${\sim}2\%$ for a higher inclination of $60^\circ$).
This albedo is broadly consistent with our expectation based on the broader flux distribution (see $\S$\ref{sec:SED}, i.e., since the dust appears to be bright in its thermal emission) and consistent with the dust albedo of other A-type stars \citep[e.g., Fomalhaut, see][]{Kalas05}. 

Although faint, we note here the tentative possibility of a scattered light asymmetry, with the dust emission more extended to the NW of the disc than the SE, most easily seen in Fig.~\ref{fig:kCrB_B6_HST}, right panel.
At the same intensity of emission, the NW extends out to ${\sim}7.0''$, whereas in the SE, this same intensity emission extends only as far as ${\sim}$5''. 
Although we do not extend analysis of this tentative asymmetry further, we suggest future observations in scattered light may better constrain this with more sensitive measurements.

\subsection{Flux distribution}
\label{sec:SED}
\begin{figure}
    \includegraphics[width=0.475\textwidth]{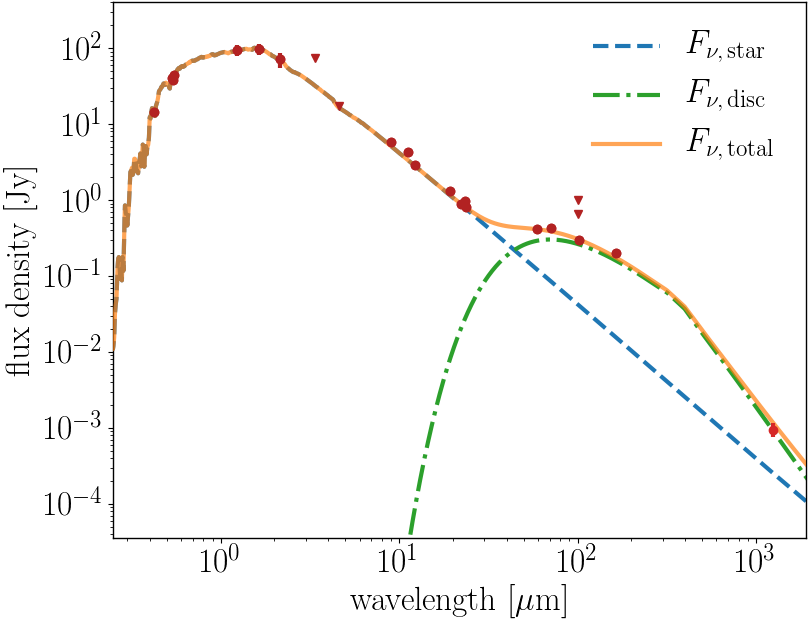}
    \caption{Flux density distribution for $\kappa\,$CrB, where the total flux measurements are dominated by emission due to the star up to ${\sim}30\,\mu$m, and by the disc emission between ${\sim}30\,\mu$m and 1.3\,mm. The disc is modelled here as a single component modified blackbody as described in $\S$\ref{sec:SED} which successfully interprets the mid-infrared and new ALMA Band~6 data. }
    \label{fig:SED}
\end{figure}

Fig.~\ref{fig:SED} shows the flux density distribution of $\kappa$\,CrB, including the flux density derived from the Band~6 ALMA data (see $\S$\ref{sec:ALMAobsAnalysis}), for which wavelengths, fluxes and magnitudes (up to K-band photometry) are given in Table~\ref{tab:SEDflux}. 
To investigate where the emission in this system originates from, we fitted the complete flux distribution with a stellar photosphere plus a `modified' blackbody, applying the methodology of \citet{Kennedy14} and \citet{Yelverton19}. 
We find that the distribution is best fitted with a single, outer, cool temperature dust component, and a star with a stellar effective temperature of $T_{\rm{eff}}{=}4850{\pm}100\,\rm{K}$, and a stellar luminosity and radius of $L{=}12.0\,L_{\odot}$, and $R{=}4.9\,R_{\odot}$ respectively (with 2\% calibration uncertainties). 
We note that this modified blackbody model is simply a normal blackbody over all wavelengths, until a specific wavelength, $\lambda_0$, whereby the flux density is then modified by a term $(\lambda/\lambda_0)^{-\beta_0}$, and that the stellar spectrum model covers the wavelength range from 0.05$\mu$m--1.0\,cm.
We further note that \citet{Gaia18} estimated these stellar parameters (using the ``Priam" and ``FLAME" algorithms) as $4750\,\rm{K}$, $12.7\,L_{\odot}$, and $5.1\,R_{\odot}$ respectively, which are all broadly in agreement with these model-fitted values. 
Error flags on the WISE1 and WISE2 data points meant that these were excluded from the fitting procedure (though these are still plotted on Fig.~\ref{fig:SED}).

The single-component model finds the disc is best-fitted with a blackbody temperature of $T_{\rm{bb}}{=}72{\pm}4\,\rm{K}$ (corresponding to an uncorrected blackbody radius of $r_{\rm{bb}}{=}51{\pm}6$\,au), a relatively faint but significant fractional luminosity of $4.9{\pm}0.5{\times}10^{-5}$, and modified blackbody parameters $\beta_0{\sim}1.6$ and $\lambda_0{\sim}400\,\mu$m.
The profile turn-over at 400\,$\mu$m can be seen clearly in Fig.~\ref{fig:SED}, though we note here that with only one data point beyond ${\sim}160\,\mu$m, $\lambda_0$ and $\beta_0$ are poorly constrained and degenerate. 
It is well known that blackbody radii measured from SEDs are smaller than the radii determined from resolved millimetre wavelength images \citep[see e.g.,][]{Pawellek15}. 
We also find this to be the case here: whilst the uncorrected blackbody radius is ${\sim}51$\,au, the peak millimetre-emission radius is ${\sim}110$\,au, and so a factor of ${\sim}2$ different.
In \citet{Pawellek15}, the ratio of these two radii measurements, $\Gamma$, is empirically quantified in equation~8 with respect to the stellar luminosity and dust composition.
For a range of compositions \citep[see Table~4 of][]{Pawellek15} and the previously determined  stellar luminosity of $L{=}12.0\,L_{\odot}$, we find a $\Gamma$ ratio of between 2.1--2.3.
Since there is good consistency between $\Gamma {\times} r_{\rm{bb}}{=}107{-}117$\,au and the ${\sim}110$\,au peak millimetre-emission radius, this demonstrates that our measurements align well with those of discs around similarly luminous stars.

We note that unresolved millimetre emission coincident with the location of the star could exist in the warmer inner regions of this system, for example, from a separate inner warm/hot dust component near to the star.
Indeed, \citet{Absil13} detected a near-infrared K-band excess at the 1 percent level, which was interpreted as due to the evolutionary stage of the star.
However, we are unable to verify the presence of a millimetre excess at 1.3\,mm co-located with the star. 
To demonstrate this, we note that our modelled stellar flux distribution has a value of 220\,$\mu$Jy at $\lambda{=}$1.3\,mm which is consistent within 1.5$\sigma$ with the measured flux at the location of the star ${\sim}191{\pm}20\,\mu$Jy as measured in $\S$\ref{sec:ALMAobsAnalysis} and shown in Fig.~\ref{fig:SED}.
Consequently, any near-infrared excess towards $\kappa\,$CrB \citep[such as that measured by][]{Absil13} must be sufficiently faint to have avoided detection at millimetre wavelengths.
We thus conclude that across the complete flux distribution of $\kappa$\,CrB, our modelling only finds significant evidence for emission due to a photospheric component and a single disc at many tens of au.

\begin{table}
    \centering
    \caption{Values for the fluxes used to produce the flux density distribution in Fig.~\ref{fig:SED}. a= \citet{2006yCat.2168....0M}, b= \citet{2015A&A...580A..23P}, c= \citet{2000A&A...355L..27H}, d= \citet{Gaia18}, e= \citet{ESA97}, f= \citet{2003tmc}, g= \citet{2010AJ....140.1868W}, h= \citet{2010A&A...514A...1I}, i= \citet{1988iras6A}, j= \citet{2010ApJS..191..212S}, k= \citet{Sibthorpe18}, l= this work. All wavelengths are provided to 3 significant figures. The calculation of the ALMA values in the lower part of this table are derived in $\S$\ref{sec:ALMAobsAnalysis}.}
    \begin{tabular}{c|c|c|c}
         \hline
         \hline
         Source & Wavelength & Magnitude & Flux \\
         & [$\mu\rm{m}$] & [mag] & [Jy] \\
         \hline
%         U-B $^{\rm{a}}$ & - & $0.873{\pm}0.005$ & - \\
%         c$_1$ $^{\rm{b}}$ & - & $0.443{\pm}0.01$ & - \\
         B$_{\rm{T}}$ $^{\rm{c}}$ & 0.420 & $6.075{\pm}0.014$ & $14.64{\pm}0.21$ \\
%         m$_1$ $^{\rm{b}}$ & - & $0.42{\pm}0.012$ & - \\
%         B-V $^{\rm{a}}$ & - & $0.996{\pm}0.005$ & - \\
%         B-Y $^{\rm{b}}$ & - & $0.615{\pm}0.008$ & - \\
         G$_{\rm{BP}}$ $^{\rm{d}}$ & 0.513 & $5.048{\pm}0.002$ & $32.5{\pm}0.3$ \\
         V$_{\rm{T}}$ $^{\rm{c}}$ & 0.532 & $4.908{\pm}0.009$ & $40.7{\pm}0.4$ \\
         H$_{\rm{P}}$ $^{\rm{e}}$ & 0.542 & $4.959{\pm}0.005$ & $38.46{\pm}0.25$ \\
         V$_{\rm{J}}$ $^{\rm{a}}$ & 0.550 & $4.812{\pm}0.009$ & $43.7{\pm}0.9$ \\
         G$_{\rm{GP}}$ $^{\rm{d}}$ & 0.642 & $4.460{\pm}0.003$& $53.0{\pm}0.5$ \\
         G$_{\rm{RP}}$ $^{\rm{d}}$ & 0.780 & $3.889{\pm}0.004$ & $71.5{\pm}0.7$ \\
         J $^{\rm{f}}$ & 1.24 & $3.035{\pm}0.184$ & $95{\pm}16$ \\
         H $^{\rm{f}}$ & 1.65 & $2.575{\pm}0.18$ & $98{\pm}16$ \\
         K$_{\rm{S}}$ $^{\rm{f}}$ & 2.16 & $2.423{\pm}0.242$ & $71{\pm}16$ \\
         \textit{WISE} W1 $^{\rm{g}}$ & 3.38 & - & $<74.7$ \\
         \textit{WISE} W2 $^{\rm{g}}$ & 4.63 & - & $<17.1$ \\
         \textit{AKARI} IRC $^{\rm{h}}$ & 8.98 & - & $5.80{\pm}0.03$ \\
         \textit{IRAS} $^{\rm{i}}$ & 11.2 & - & $4.34{\pm}0.18$ \\
         \textit{WISE} W3 $^{\rm{g}}$ & 12.3 & - & $2.93{\pm}0.12$ \\
         \textit{AKARI} IRC $^{\rm{h}}$ & 19.2 & - & $1.396{\pm}0.021$ \\
         \textit{WISE} W4 $^{\rm{g}}$ & 22.3 & - & $0.88{\pm}0.05$ \\
         \textit{IRAS} $^{\rm{i}}$ & 23.3 & - & $1.02{\pm}0.04$ \\
         \textit{MIPS} $^{\rm{j}}$ & 23.7 & - & $0.800{\pm}0.008$ \\
         \textit{IRAS} $^{\rm{i}}$ & 59.4 & - & $0.41{\pm}0.03$ \\
         \textit{MIPS} $^{\rm{j}}$ & 71.4 &  - &$0.425{\pm}0.023$ \\
         \textit{IRAS} $^{\rm{i}}$ & 100 &  - &$0.65{\pm}0.14$ \\
         \textit{PACS} $^{\rm{k}}$ & 101 &  - &$0.2959{\pm}0.0016$ \\
         \textit{PACS} $^{\rm{k}}$ & 164 &  - &$0.202{\pm}0.003$ \\
         \hline
         ALMA B6 $^{\rm{l}}$ & 1340 & - & $0.00096{\pm}0.00010$ \\ 
         \hline
    \end{tabular}
    \label{tab:SEDflux}
\end{table}

\begin{table*}
    \centering
    \caption{Posterior best-fit values for the two models SGR and SPL. Dashes here denote parameters that were either not used given the different model parametrisations, or in the case of $r_2$ where only a lower limit could be derived. Numbers in brackets indicate the ${\pm}1\,\sigma$ error as measured by the posterior distributions of the mcmc fits.}
    \begin{tabular}{l|c|c|c|c|c|c|c|c|c|c|c|c}
         \hline
         \hline
          & $M_{\rm{dust}}$ & $r_1$ & $i$ & $\rm{PA}$ & $F_{\star\,\rm{B6}}$ & $x_{\rm{off}}$ & $y_{\rm{off}}$ &  $p_{\rm{SPL}}$& $r_2$ & $\sigma_{\rm{SGR}}$ \\
         &[$M_{\oplus}$]&[au]&[deg]&[deg]&[$\mu$Jy]&[$''$]&[$''$]&[au]&&[au] \\
         \hline
%         SNR & 0.0094 &115 &61.7 &145.1 &0.196 &0.08 &-0.20 & -& -&-\\
%         & (0.0010)& (4)& (3.2)& (3.6)& (0.013)& (0.04)& (0.07)& -& -&-\\
         \hline
         SPL &0.0158 &93 &62.3 &148 &195 &0.07 &-0.20 & -2.5 & ${>}151$&-\\
         & (0.0026)& (8)& (3.4)& (5)& (13)& (0.04)& (0.07)& (1.2) &-&- \\
         \hline
%         DNR &0.0140 &112 &64.7 &147.9 &0.193 &0.09 &-0.21 & -& 191&0.54&- \\
%         & (0.0016)& (4)& (2.8)& (3.7)& (0.013)& (0.04)& (0.07)& - &(17)& (0.14)&-\\
%         \hline
         SGR &0.0170 &131 &61.1 &151 &193 &0.07 &-0.20 & -& -&98 \\
         & (0.0028)& (9)& (3.8)& (5)& (13)& (0.04)& (0.07)& - & - &(19)\\
         \hline
    \end{tabular}
    \label{tab:bestFitVals}
\end{table*}

\section{Modelling $\kappa$\,CrB's Dust Observations}
\label{sec:modelling}
Following our observational analysis in $\S$\ref{sec:dataAnalysis}, here we assess the morphology of $\kappa$\,CrB's debris disc with parametric models.
We start by discussing our Band~6 model setup in $\S$\ref{sec:modellingsetup}, their results and implications in $\S$\ref{sec:modellingresults}, explore how this relates to a possible model of the dust emission in scattered light in $\S$\ref{sec:modelHST}, and summarise our key modelling findings in $\S$\ref{sec:modellingsum}.

\subsection{Modelling the millimetre emission}
\subsubsection{Millimetre model setup}
\label{sec:modellingsetup}
The models described in $\S$\ref{sec:obsALMAsum} share a common set of 7 free parameters: dust mass ($M_{\rm{dust}}$), main belt radius ($r_{\rm{1}}$), inclination ($i$), position angle ($\rm{PA}$), photospheric Band~6 flux ($F_{\star,\,\rm{B6}}$), and the Band~6 measurement set phase centre offsets in RA and Dec ($x_{\rm{off}}$, $y_{\rm{off}}$ respectively). 
We modelled the system with two belt distributions, `model SPL', defined by an inner edge ($r_1$), outer edge ($r_2$), and power-law profile, with an index ($p_{\rm{SPL}}$) for a total of 9 free parameters, and `model SGR', defined by a peak emission radius ($r_1$) and Gaussian width ($\sigma_{\rm{SGR}}$; the standard deviation of the Gaussian distribution) for a total of 8 free parameters. 
Although we measured the Band~6 emission coincident with the star as ${\sim} 191$\,$\mu$Jy in $\S$\ref{sec:SED}, since stellar emission in the millimetre can be poorly constrained, we leave the flux located near the star ($F_{\star,\,\rm{B6}}$) as a free parameter.
The models have the same vertical Gaussian density distribution, defined by the scale height, $h{=}H/r$, where $H$ is the height of dust above the midplane at a radius $r$ \citep[][]{Marino16}. 
Given our observations only have a low signal-to-noise and the disc is only moderately inclined, we fix $h{=}0.05$ \citep[consistent with the expected range of $h{=}$0.02-0.12 for debris discs][]{Hughes18}, and further note that by investigating model fits with $h$ left as a free parameter we always yielded unconstrained values.
Given the lack of any observed asymmetries in our images, we do not model particle eccentricities.
Additionally, all these models have identical (and fixed) minimum and maximum grain radii of $a_{\rm{min}}{=}1.0\,\mu \rm{m}$ and $a_{\rm{max}}{=}1$ $\rm{cm}$.
These are set at the lower-limit by the approximate blow out size, which is just a few microns for astrosilicate-dominated grains around a star with $1.8\,M_\odot$ and $12\,L_\odot$ \citep[see e.g.,][equation 14; $D_{\rm{blowout}}=0.8(L_\star/M_\star)(2700/\rho)$]{Wyatt08}, and at the upper-limit by neglecting emission from grains a factor of 10 larger than the Band~6 wavelength of 1.3\,mm.
We further assume a dust grain density of $\rho=2.7$\,$\rm{g}$\,$\rm{cm}^{-3}$, a size distribution with power-law exponent $\alpha{=}3.5$ \citep{Dohnanyi69}, and a weighted mean dust opacity of $\kappa{=}1.88\,\rm{cm}^2\,\rm{g}^{-1}$, based on a mix of compositions with mass fractions of 15\% water-ice, 70\% astrosilicates and 15\% amorphous carbon \citep[see both][]{Li98, Draine03}. 
We note this composition was likewise used to model the ${>}$Gyr old debris discs towards HD\,107146 and HD\,10647 \citep[see][respectively]{Marino18,Lovell21c}.
Since we do not investigate the material composition or size distribution, $a_{\rm{min}}$, $a_{\rm{max}}$, $\rho$, $\alpha$ and $\kappa$ remain fixed throughout.

\begin{figure}
    \includegraphics[width=1.0\columnwidth]{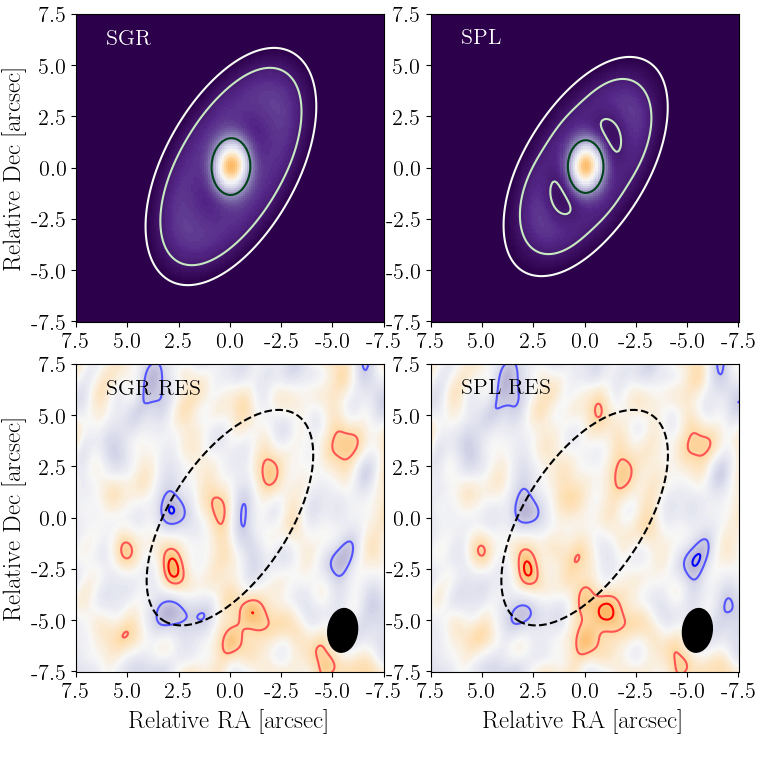}
    \caption{Best-fit parameter model and residual images (best-fit model subtracted from the data) for both Band~6 models investigated. 
    Upper plot contours (model images) are +3, 5 and 7$\,\sigma$, and lower plots (residual images) are ${\pm}2$ and 3$\,\sigma$ (red positive, blue negative). In both residual plots the beam is shown in black in the lower-right of the panels, and a circle with radius 180\,au inclined to $60\degr$ is shown with a black dashed loop.}
    \label{fig:kCrB_modsResids}
\end{figure}

\subsubsection{Millimetre modelling results}
\label{sec:modellingresults}
We used the \textit{RADMC-3D} package \citep{Dullemond12} with the mcmc parameter estimator, $\it{emcee}$ \citep{Goodman10, FM13} to compute the dust temperature of our disc models.
These dust temperatures were used to create 2D images at $1.3\,$mm and subsequently used to compute model visibilities at identical uv-baselines to our ALMA observations.
This was done identically to \citet{Lovell21c}, using the tools developed in \citet{Marino18} and \citet{Lovell21c}, where a complete discussion of model fitting is presented.

All model best-fit results are reported in Table~\ref{tab:bestFitVals}, with their corresponding model and residual images shown in Fig.~\ref{fig:kCrB_modsResids}.
We start by noting that the four parameters determined in $\S$\ref{sec:ALMAobsAnalysis}, i.e., the main belt radius ($r_1$), position angle $\rm{PA}$, inclination $i$ and Band~6 stellar flux $F_{\star\,\rm{B6}}$ are all found to within $3\,\sigma$ of both best fit Band~6 model parameter sets.
Note that whilst the power-law model SPL has a sharp inner edge found to be at $93{\pm}8\,$au, given this model is broad, when this is imaged and convolved the same ALMA beam as our measurements, this has a peak \textit{emission} radius at ${\sim}110\,$au.
These models also suggest there is ${\sim}0.016\,M_\oplus$ of dust which will be dominated by the ${\sim}$centimetre-sized grains in the model. 
We discuss the implications of this dust mass further in $\S$\ref{sec:discussionBeltMass}.
In addition, our modelling suggests there may be a phase centre offset given all $x_{\rm{off}}$ and $y_{\rm{off}}$ parameters are ${\gtrsim}1{-}2\,\sigma$ different to 0, however since there are no situations in which any of these exceed $3\,\sigma$ (in any single axis) we do not discuss this any further.
Finally, we note that in both residual images, a faint, unresolved clump is present at relative RA, Dec position of [-1.0, -5.0], with a flux density $30{-}40\,\mu$Jy, consistent with the emission of the same clump in Fig.~\ref{fig:kCrB_B6_HST} if due to an unresolved point source.
Since this is present in the residuals of both images, our model fits have likely not been strongly influenced by the presence of such a clump.
We discuss the possible origins of this clump emission further in $\S$\ref{sec:discussionClump}.

\subsubsection{A broad debris disc}
Deriving best-fit parameter estimations provides us with useful constraints on the physical morphology of the debris disc, however our modelling was primarily conducted to test whether or not our data provided evidence that the underlying planetesimal belt is extended or narrow.
To quantify this, by fitting models that can measure either the width of the emission (if fitted as a Gaussian) or the extent of the emission from the distance between the inner and outer edges (if fitted as a power-law), both our models show evidence that the emission is indeed broad.
Discussing these models in turn, we first find that the Gaussian disc attempts to fit an extremely broad disc with a radially symmetric profile with a width (standard deviation) of $\sigma_{\rm{SGR}}=98{\pm}19\,$au. %, i.e., if the emission arose from a narrow distribution, then this value would have either been found to be much lower, or unconstrained.
%Whilst this model suggests a wide belt is needed to fit the data, we find that this model is not as good at representing the data as model SPL.
Given the width of model SGR is so broad (symmetric about ${\sim}131$\,au, with a width ${\sim}98$\,au) this simultaneously attempts to fit the inner regions where emission is likely dominated by the star, and the outer regions where emission may be entirely due to noise, and thus can be seen to over-subtract and under-subtract emission in various regions of the residual image (see Fig.~\ref{fig:kCrB_modsResids}).
Whilst this model supports the hypothesis that the disc is broad, it may not be a good parametrisation of this system. 
We quantify this further in $\S$\ref{sec:modelcompstats} by comparing models SGR and SPL statistically.

In contrast, the model describing the belt with a decaying power law (SPL) provides a more promising interpretation, whilst also showing the emission to arise from a broad disc.
We quantify this breadth based on the modelled disc outer edge ($r_2$) which tends to unconstrained high values from which we can set a lower limit. 
For this reason, we quote $r_2{>}151\,$au in Table~\ref{tab:bestFitVals}, based on where ${>}99.7\,\%$ of all $r_2$ values finished in their mcmc chains.
Comparing with the well constrained inner edge of $93{\pm}8$\,au, this confirms that the outer edge is indeed significantly distant from the inner edge, and thus the disc has a broad extent.
We constrain this extent by finding the difference between the mcmc chains of the fitted inner and outer belt radii and find that over $99.7\,\%$ of these are separated by 50.1\,au (comparable with the resolution scale of our imaging, thus at least partially resolved), and assess this as a strict, significant lower limit on the belt extent. 
Based on the peak emission radius (${\sim}$110\,au), this model would therefore place a fractional width on this disc of ${>}46\%$. 
We note here that if we instead use the 50th-percentiles in the mcmc chains to find the median extent (instead of the $99.7$th-percentile) we measure a much broader extent of 135\,au (since the median outer edge radius is at a wider value of 232\,au).
As such the disc could well be much broader than the 50.1\,au lower limit, which higher signal-to-noise observations may measure.
%Higher signal-to-noise observations may therefore find this disc is indeed much broader.

\begin{figure}
    \includegraphics[width=1.0\columnwidth]{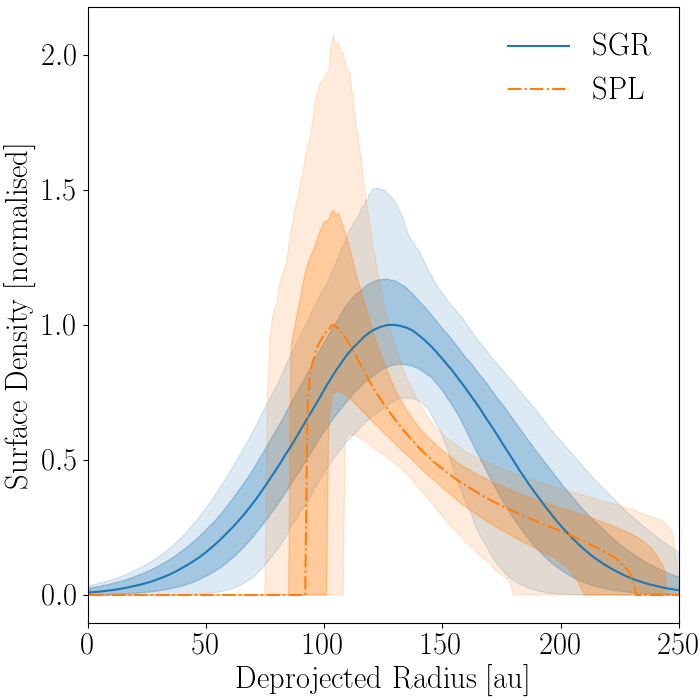}
    \caption{Surface density plots for the two models. These are plotted based on a random sample of 5000 mcmc chains (after burn-in), and show the 50th percentile plots, ${\pm}$1 and 2$\,\sigma$.}
    \label{fig:modelsSurfDen}
\end{figure}

To demonstrate further the broad extent of the underlying models, we show in Fig.~\ref{fig:modelsSurfDen} the \textit{surface density} profiles of the two models, based on the 50th percentile of the mcmc chains after burn-in, along with their corresponding ${\pm}1\sigma$ and ${\pm}2\,\sigma$ results.%, i.e., those falling within the 15.9-84.2\% and 2.3-97.7\% bins respectively.
Both models show that their 50th-percentile surface densities exceed 0 over a broad range of radii, i.e., that broad radial ranges are required for the modelled dust in order to interpret the Band~6 observations.

\begin{table}
    \centering
    \caption{Statistical model results for both tested model types based on their best-fit parameters, with reference to the SGR model.}
    \begin{tabular}{l|c|c|c}
         \hline
         \hline
         Model & $N_{\rm{params}}$ & $\Delta \chi^2$ & $\Delta \rm{BIC}$ \\
         \hline
         SGR & 8 &  - & - \\
         SPL & 9 & -4.7 & +11.6 \\
         \hline
    \end{tabular}
    \label{tab:chi2BIC}
\end{table}

\begin{figure*}
    \includegraphics[width=1.0\textwidth]{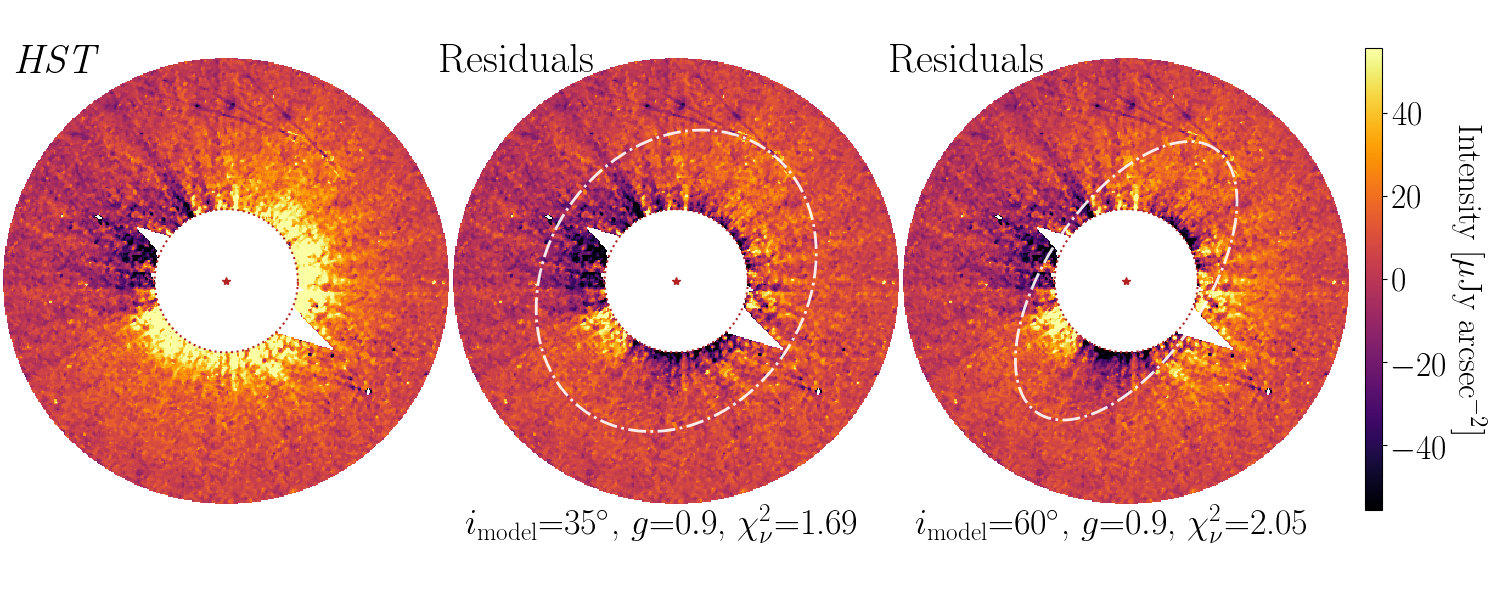}
    \caption{Scattered light images used to model the HST emission, in all three panels north is up, and east is left. Left: HST image (as per Fig.~\ref{fig:kCrB_B6_HST}). Centre and Right: residual images (data-model) for disc models with Henyey-Greenstein scattering factors of 0.9, power-law indices of $\gamma{=}0.0$, and either inclinations of $i{=}35^\circ$ or $i{=}60^\circ$ respectively. In the $i{=}35^\circ$ model, $\chi^2_\nu{=}1.69$ (the lowest of all models), whereas in the $i{=}60^\circ$ model, $\chi^2_\nu{=}2.05$, i.e., a less good fit. Also plotted are white dash-dot ellipses demonstrating 200\,au rings inclined at either 35$^\circ$ (centre) or 60$^\circ$ (right). }
    \label{fig:HSTmodelresid}
\end{figure*}

\subsubsection{Millimetre model statistical comparison}
\label{sec:modelcompstats}
Although qualitatively model SPL appears to be a better representation of the data, we additionally compare models SPL and SGR statistically. 
We do so by assessing their respective $\chi^2$ and $\rm{BIC}$ values, with the former assessing the goodness-of-fit of the models, and the latter being a comparative assessment based on the size of our interferometric data sets, and number of parameters in our model, i.e., $\rm{BIC} = \chi^2 + N_{\rm{par}}\ln{N_{\rm{dat}}}$ \citep[see][for which a difference in BIC values of 10 is considered strong evidence to support a given hypotheses, e.g., that one model is more statistically significant than another]{Schwarz78,Kass95}.
For our Band~6 data, there are $2\times N_{\rm{Vis}} = N_{\rm{Dat}}=12,104,328$ independent data points, accounting for the real and imaginary components of the interferometric visibilities. 
We report in Table~\ref{tab:chi2BIC} these relative values, and show with reference to the SGR model that whilst the additional parameter included in the SPL model improves the fit to the data (i.e., model SPL has a lower $\chi^2$), this may not be significant (i.e., model SGR has a lower $\rm{BIC}$).
Whilst this may suggest that the increase in the number of parameters from 8 to 9 between models SGR and SPL is not favoured, \citet{Lovell21c} demonstrated that statistical comparisons of disc model fits to low signal-to-noise interferometric data sets can be significantly skewed by large $N_{\rm{Dat}}$.
Consequently, as $\Delta \rm{BIC}$ may not be a meaningful comparative diagnostic, this would suggest that model SPL is statistically favourable of the two models.

\subsubsection{Alternative models?}
Whilst the data can be fitted with a smooth wide disc, we cannot rule out the possibility that there is a gap or narrow radial sub-structure in the disc that gives rise to the observed emission structure.
Albeit with less significant fits in additional tests, i.e., poorer $\chi^2$ and $\rm{BIC}$ values, and visibly poorer residuals, we found that the data can nevertheless be modelled by \textit{two} narrow belts.
Such a scenario comprising multiple planetesimal belts would not be unprecedented as it has been found that gaps in wide debris discs may be common \citep[see, e.g.,][]{Marino20}.
However, given the low signal-to-noise, the resolution of our observations, and that such a model was indeed a poorer representation, we do not assess this possibility any further in this work.

\subsection{Modelling the scattered light emission}
\label{sec:modelHST}
Whilst the emission in the millimetre traces well the large planetesimals within a disc, due to stellar radiation forces affecting much smaller grains, tracing the parent bodies with small particles is much more challenging.
This means that the observed morphologies of discs in scattered light and in the millimetre can be substantially different, even if their dust formed in the same parent belts \citep[for example, see Beta Pic as imaged in scattered light with HST, and in the millimetre with ALMA, see][respectively]{Golimowski06,Dent14}.
Nevertheless, constraints on disc morphologies such as their position angles and inclination can be obtained from scattered light imaging, and comparisons with discs as observed in other wavelengths probe different disc processes. 
As such, we outline here our investigation into the morphology of the scattered light emission for which we model the HST data in order to compare this with the emission observed in the millimetre images.

We start by noting that our scattered light HST model is simply a variant of the broad single power-law millimetre model (model `SPL') described in $\S$\ref{sec:modellingsetup}, but instead set up to produce images of scattered light emission at 0.6\,$\mu$m, with a range of anisotropic scattering factors ($g$) based on the Henyey-Greenstein (HG) scattering function \citep{HG41}. 
In this investigation, we fixed the inner and outer edges of the disc to 73\,au and 280\,au respectively (based on the peak brightness of the HST observations, and the outer edge at which scattered light emission was observed in $\S$\ref{sec:HSTobsAnalysis}), the position angle to 148$^{\circ}$, and produced a suite of models, based on variations of the disc inclination (ranging from 25 to 60$^\circ$ in 5$^\circ$ increments), scattering factors (ranging from 0.1 to 0.9 in 0.1 increments) and power-law indices (ranging from -2 to 2 in 0.5 increments).
To find the best-fit of the model suite, we introduced a dummy pre-factor which each model was multiplied by before the data was subtracted (used to scale the emission brightness to maximise the goodness-of-fit, which was also varied in order to minimise the $\chi^2$ of each model).
To avoid our goodness-of-fit estimates being driven by emission far from the star and by noisy inner regions, a null map was produced based on the data and convolved with all images, such that models contained identical imaging artefacts as the raw data. 
A noise estimate from the raw data was obtained from emission unaffected by real scattered light emission, i.e., beyond 280\,au, with which the reduced $\chi^2$, $\chi^2_\nu$, of each residual image (HST data minus model) could then be measured. 
We note here that our null model with zero emission everywhere has $\chi^2_{\rm{\nu,\,null}}=3.75$, whereas the suite of models ranged in their measured $\chi^2_{\nu}$ values between 1.69--2.70.
Further, by simulating grids of noise, we estimate the uncertainty in measured $\chi^2_{\nu}$ values as ${\sim}$0.10.

We show in the left panel of Fig.~\ref{fig:HSTmodelresid} the HST data (unchanged from Fig.~\ref{fig:kCrB_B6_HST}), in the centre panel the best-fit model residuals, and in the right panel the best-fit model residuals for a disc inclined to 60$^{\circ}$ (consistent with that measured in the millimetre).
We note that the overall best-fit model has a $\chi^2_{\nu} {=} 1.69$ and the best 60$^{\circ}$ inclined model has a $\chi^2_{\nu} {=} 2.05$.
This investigation found generally that to minimise $\chi^2$, models required low inclinations in the range $i{=}30{-}40^\circ$, highly anisotropic HG scattering factors in the range $g{\sim}0.8{-}0.9$, and relatively flat surface density power-law indices in the range $\gamma=-0.5{-}0.5$ (consistent with our imaging analysis earlier, see $\S$\ref{sec:HSTobsAnalysis}). 
We note here that 1) variations in the disc position angle (between 142--152$\degr$) did not yield any significant improvement in the measured $\chi^2_{\nu}$ values, and 2) increasing the value of the scale height, $h$ (to either 10\%, 15\% or 20\%) did not affect the preferred inclinations of best-fitting model discs.
The second of these points suggests that there is no measurable dependency between the scale height and the disc inclination.
However, we do not rule out that different model parametrisations could yield different constraints on these parameters, and in particular different phase scattering functions, which we briefly investigate later in this section, and discuss further in \S\ref{sec:discussionBeltInc}.

Discussing these two residual images in turn, in the case of the best-fit (centre panel) there remains emission in the region between 73-280\,au, however the broad azimuthal region covering the south-east to the north-west is reasonably well modelled. % over the regions in which the HST scattered light emission is strongest.
In addition, in the north-east where the HST image artefacts are significant (but the absolute scattered light emission is lower), the model accounts for this with its high scattering factor which naturally lowers the emission brightness on the far-side of the disc.
Therefore, whilst this minimal $\chi^2_{\nu}=1.69$ suggests a better fit to the data can be found, this model does at least broadly account for the morphology of the scattered light emission, and is statistically favoured to the best-fit model with $i{=}60^\circ$ inclination.
For the purposes of this work we deem this $i=35^\circ$ best-fit model to be sufficiently accurate for comparative purposes. 
However, detailed modelling of $\kappa\,$CrB's scattered light emission could be conducted in the future to investigate this in greater detail and understand the uncertainty distribution of best-fit parameters.
We explore some of the details that require further investigation in $\S$\ref{sec:discussionBeltInc}.

In the case of a model identical except for having a more inclined $i{=}60^\circ$ disc, the measured $\chi^2_{\nu}$ is significantly higher (2.05), and it can be seen in the image residuals (right panel, Fig.~\ref{fig:HSTmodelresid}) that emission in the azimuthal region between the south-east and north-west is not as well modelled: there is brighter excess emission in the south-west, and a deeper subtraction along the disc major axis in the south-east and north-west.
This demonstrates that the model fits in scattered light favour an inclination that is shallower than the inclination measured and modelled in the millimetre \citep[and those in the far-infrared, see][]{Bonsor13}.
By investigating a range of different HG scattering factors, inclinations, and power-law indices, we have shown that this is a trend seen in all disc models: with highly anisotropic scattering factors, relatively flat power-law indices, and low ${\sim}30{-}40^\circ$ inclinations favoured.

To test whether the millimetre-scattered light inclination discrepancy arises from the HG scattering parametrisation, we further investigated whether a more inclined scattered light model could fit the observations by considering spherical grains and Mie-theory, consistent with the models presented in \citet{Pawellek19}.
Mie-theory was investigated since strong forward scattering is expected \citep[see the Appendix of][]{Pawellek19}.
In doing so, we again found that the scattered light models prefer low inclination values between 30-40$^{\circ}$, consistent with the modelling approach presented above, and still discrepant with that observed with ALMA and Herschel.
We explore the implications for this inclination discrepancy further in $\S$\ref{sec:discussionBeltInc}.

\subsection{Summary of modelling}
\label{sec:modellingsum}
To summarise, by comparing the outputs of parametrised models fitted to the ALMA Band~6 visibilities, we find that $\kappa\,$CrB's debris disc most likely arises from a radially broad distribution of planetesimals, which we constrained as having an extent ${>}50\,$au.
In particular, we showed that whilst radial power-law and Gaussian models can fit the data, the power-law model is favourable. 
However, given the disc emission could have been broadened due to the presence of unresolved multiple planetesimal belts, future deep millimetre observations of the $\kappa\,$CrB system could be undertaken to examine alternative models. 

Additionally, we produced simple parametric models of the scattered light emission to investigate the dust emission seen by HST.
We showed with that the models that best explain the observed dust emission require the emission to be i) highly anisotropic with Henyey-Greenstein scattering factors between $g=0.8-0.9$, ii) to have a relatively flat low power-law index between $\gamma=-0.5{-}0.5$, and iii) to have low inclination angles between $i=30{-}40^\circ$.
These models come with a range of uncertainties and are not able to fully interpret the HST emission, thus could be greatly improved with detailed modelling and new scattered light data.
Nevertheless, we have shown the disc to have a less inclined morphology in scattered light to that in the millimetre. 
We discuss the implications of the scattered light and millimetre modelling further in $\S$\ref{sec:discussion}.

\begin{figure}
    \includegraphics[width=1.0\columnwidth]{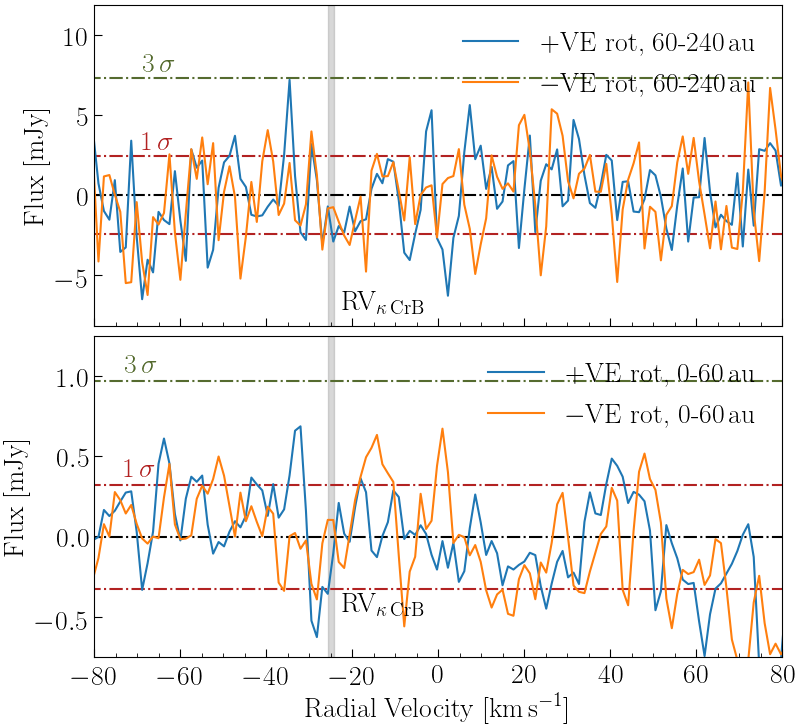}
    \caption{CO spectral data integrated between a deprojected distances from the star, where a significant detection would coincide with the stellar radial velocity marked in grey. Top: covering a range between 60-240\,au, and bottom: covering a range from 0-60\,au. }
    \label{fig:kCrB_CORV}
\end{figure}

\section{CO J=2-1 Analysis}
\label{sec:COflux}
Alongside measurements of the dust continuum, the ALMA Band~6 observations as introduced in $\S$\ref{sec:dataRed} included higher spectral resolution measurements over the CO J=2-1 spectral line frequency.
Therefore, we produced a subset of the \textit{CASA} measurement set containing only data covering this spectral line with rest frequency $f_{\rm{CO}}{=}230.538$\,GHz. 
To remove continuum emission from any $\rm{CO}$ signal, we used the \textit{CASA} tool $\rm{uvcontsub}$ with a fit order of 1 \textit{external} to the region defined by $f_{\rm{CO}} {\pm} \Delta f$, where $\Delta f$ was set to a width $30\,$MHz, avoiding fitting to channels where $\rm{CO}$ could be present. 
We then produced a data cube in the barycentric reference frame using the \textit{CASA} $\rm{tclean}$ algorithm with zero clean iterations, since there was no significant emission per beam present in any channel. 

We note that the relative line-of-sight velocity of any ${\rm{CO}}$ gas emission (if present) would however fall across multiple data channels due to the variation of the observed radial velocity around the disc. 
This can however be accounted for since the radial velocity of the star, the disc morphology and stellar mass are all known, for which $v_{\rm{rad}}{=}-24.98{\pm}0.14\,\rm{kms^{-1}}$ \citep[see][]{Gaia18}. 
Given the disc is broad, any gas that is co-orbiting with the disc will span a range of velocities and thus fall over a range of channels.
To measure the line profiles from our CO data cube, we therefore applied the spectro-spatial filtering method of \citet{Matra15,Matra17} by producing a Keplerian mask to shift pixels between cube channels based on a stellar mass of 1.8\,$M_\odot$ \citep[see][]{Johnson08} and the same disc position angle, inclination as derived in $\S$\ref{sec:ALMAobsAnalysis} in both of the possible disc rotation directions, (either clockwise and anti-clockwise) and integrated all emission (in both rotation directions separately) between projected radii of either 0-60\,au or 60-240\,au from the star.
These two distances cover the two regions dominated in the continuum by either i) the star from 0-60\,au, or ii) by the disc from 60-240\,au.
Both spectra are shown as a function of radial velocity in Fig.~\ref{fig:kCrB_CORV}, which show that no significant CO J=2-1 line emission is present towards $\kappa$\,CrB, in either of the spatial regions, and in either of the rotation directions.

We found the rms of the profiles (in the two integrated regions) across all velocities, and measured $\rm{rms_{inner}{=}0.32\,\rm{mJy}}$ in the inner 0-60\,au region (averaged in both clockwise and anti-clockwise directions), and $\rm{rms_{outer}{=}2.5\,\rm{mJy}}$ in the outer 60-240\,au region (averaged in both clockwise and anti-clockwise directions). 
We thus define a $3\,\sigma$ upper bound CO line peak flux (based on the 60-240\,au region) of $F_{\rm{CO,3\,\sigma}}{=}7.5\,\rm{mJy}$. 
Although our channel spacing equates to a velocity width, $\Delta v {=} 1.269\,\rm{km\,s}^{-1}$, the \textit{effective} spectral bandwidth is ${\sim}2.667{\times}$ larger than this (since adjacent channels in ALMA data are not fully independent from each other\footnote{A more complete discussion of this is provided in \url{https://safe.nrao.edu/wiki/pub/Main/ALMAWindowFunctions/Note_on_Spectral_Response.pdf}}).
Therefore, we place a $3\,\sigma$ limit on the integrated CO flux as $\delta F_{\rm{CO},3\,\sigma} {=} 2.667 \times \Delta v \times \rm{rms} {=} 25.4\,\rm{mJy\,kms^{-1}}$. 
Since $\kappa\,$CrB has more recently been interpreted as having a lower stellar mass \citep[i.e., $1.32\,M_\odot$, see][and also Fig.~\ref{fig:HRdiag}]{White20}, we also checked the robustness of this CO flux limit by re-measuring this flux with a lower stellar mass of $M_\star{=}1.32\,M_\odot$, which also yielded a consistent non-detection. % limit, i.e., both spectral masks from the two stellar mass values yielded consistent non-detections and upper-limit integrated flux measurements of the CO J=2-1 spectral line.
We discuss the CO J=2-1 observations further in $\S$\ref{sec:discussionCO}, where we use our CO flux upper limit to constrain the total CO gas mass, and the possible origins of CO.

\section{The $\kappa$\,CrB Planetary System}
\label{sec:planetArch}
Although there is a known radial velocity (RV) planet, a low-eccentricity exo-Jupiter with a ${\sim}1250$ day orbit \citep[see][and our introduction]{Johnson08}, long-term RV measurements of $\kappa$\,CrB over the space of 8 years found this to have a gradient in its RV trend that may have been the result of acceleration from an outer massive body \citep[i.e., a large planet or brown dwarf, see][]{Bonsor13}.
If this RV trend is indeed due to an outer body, to evade being detected (for a given orbital radius) it would need to be sufficiently faint \citep[i.e., to have been undetected by the adaptive optics of Keck II observations, see][for a complete discussion]{Bonsor13}, and have sufficiently low-mass \citep[i.e., to avoid producing a significant proper motion anomaly in its Gaia eDR3-Hipparcos linear motion solution, see][for a full discussion]{Brandt21}.
Since the publication of \citet{Bonsor13}, the ${\sim}$8 years of RV data (2004-2011) has expanded to ${\sim}$17 years (2004-2021) with new APF observations \citep[Automated Planet Finder telescope, at the Lick Observatory, see][]{Vogt14,Radovan14} and HIRES observations \citep[the high-resolution echelle spectrometer, at the Keck Observatory, see][]{Vogt94}. 
Given the doubled baseline of observations, we updated the constraints on the putative outer body using the $\rm{RadVel}$ package \citep{Fulton18}. 

$\rm{RadVel}$ allows users to define the orbital parameters for N bodies, for which we investigated the goodness-of-fit of models with N=0, 1 or 2 planets. 
In Table~\ref{tab:RVparams} we report our model comparison results (assuming a stellar mass of $M_\star=1.8{\pm}0.3\,M_\odot$ and letting the jitter parameter vary freely) and start by noting that the N=0 planet model is strongly ruled out, but that N=1 and N=2 planet models are more comparable. 
We note that the parameters of the inner planet in both N=2 and N=1 models remain strongly consistent with the previously confirmed exo-Jupiter planet in the literature \citep[e.g.,][]{Johnson08}, and further, the inner planet of the N=2 planet model has consistent values with the single derived planet in the N=1 fit, as shown in Table~\ref{tab:bestfitplanets}.
Whilst this model comparison investigation suggests that a 2-planet configuration is statistically favoured, we argue against such a configuration on dynamical grounds.
In the modelled scenario, whilst the inner planet remains in the same location, the model derives the outer planet mass as $M_{\rm{pl\,c}}\sin{i}{=}0.39{\pm}0.10\,M_{\rm{Jup}}$, the semimajor axis as $a{=}6.0{\pm}0.4$\,au, and the eccentricity as $e{=}0.78{\pm}0.13$.
Such an eccentricity and orbital radius result in this body having a periastron at ${\sim}1.3\,$au, i.e., inside the 2.8\,au orbit of the exo-Jupiter, $\kappa$\,CrB~b, which we deem an unstable configuration over the 2.5\,Gyr age of the system.
From this, we conclude that whilst the RV baseline has doubled from 8 to 17 years, $\rm{RadVel}$ is unable to determine the orbit of a stable N=2 planetary configuration from the RV data.
We show in Appendix~\ref{sec:AppendixA} Fig.~\ref{fig:appRadVel} the radial velocity plots of the N=2 parameter model.

\begin{table}
    \centering
    \caption{Comparison statistics for the N=2 model explored by $\rm{RadVel}$.}
    \begin{tabular}{l|c|c|c}
         \hline
         \hline
         Statistics & Planets: 0 & Planets: 1 & Planets: 2 \\
         \hline
         $N_{\rm{data}}$ & 236 & 236 & 236 \\
         $N_{\rm{RV\,params}}$ & 6 & 12 & 18 \\
         $\sigma_{\rm{RV}}$ & 13.7 & 5.45 & 5.11 \\
         $\ln{L}$ & -944.2 & -735.8 & -717.9\\
         $\rm{BIC}$ & 1911.1 & 1527.1 & 1524.0 \\
         \hline
    \end{tabular}
    \label{tab:RVparams}
\end{table}

\begin{table}
    \centering
    \caption{N=1 and N=2 RV planet model best-fit parameters, and orbital derived parameters.}
    \begin{tabular}{l|c|c|c}
         \hline
         \hline
         Fitted parameters & N=1 value & N=2 value & Units \\
         \hline
         Period ($P_{\rm{b}}$)& 1248.4$^{+1.5}_{-1.6}$ & 1247.9$^{+4.9}_{-9.4}$& days \\
         Eccentricity ($e_{\rm{b}}$)& 0.083${\pm}0.009$ & 0.059${\pm}0.033$& - \\         
         Arg. of pericentre ($\omega_{\rm{b}}$) & $2.13{\pm}0.13$ & $1.7{\pm}0.6$ & radians \\
         Semi-amplitude ($K_{\rm{b}}$) & $23.5{\pm}0.3$ & $22.8{\pm}0.9$ & m\,s$^{-1}$ \\
         RV velocity trend ($\dot v$) & $0.73{\pm}0.07$ & $0.83{\pm}0.27$ & m\,s$^{-1}$\,yr$^{-1}$ \\
         Period ($P_{\rm{c}}$)& - & 3936$^{+350}_{-83}$& days \\
         Eccentricity ($e_{\rm{c}}$) & - & 0.78${\pm}0.13$& - \\         
         Arg. of pericentre ($\omega_{\rm{c}}$) & - & $1.17^{+0.36}_{-0.32}$ & radians \\
         Semi-amplitude ($K_{\rm{c}}$) & - & $5.5^{1.9}_{-1.3}$ & m\,s$^{-1}$ \\
         \hline
         Derived Parameters & & & \\
         \hline
         Mass $\sin{i}$ ($M_{\rm{b}}$) & $1.84{\pm}0.14$ & $1.78{\pm}0.21$ &  $M_{\rm{Jup}}$ \\
         Semi-major axis ($a_{\rm{b}}$) & $2.76{\pm}0.11$ & $2.76{\pm}0.16$ & au \\
         Mass $\sin{i}$ ($M_{\rm{c}}$) & - & $0.39{\pm}0.10$ &  $M_{\rm{Jup}}$ \\
         Semi-major axis ($a_{\rm{c}}$) & - & $6.0{\pm}0.4$ & au \\
         \hline
    \end{tabular}
    \label{tab:bestfitplanets}
\end{table}

Nevertheless the N=2 and N=1 planet best-fit models still retain a significant RV acceleration trend (which in the case of the N=2 model is $\dot v = 0.83{\pm}0.27$\,ms$^{-1}$\,yr$^{-1}$, as shown in Table~\ref{tab:bestfitplanets}).
Although this acceleration term may not necessarily be due to an outer planet, we cannot rule out this acceleration term as originating from an outer body that cannot be well-fitted by $\rm{RadVel}$. 
We can constrain the orbit and mass of such a body in a number of ways.
Firstly, we estimate the minimum orbital radius as ${\sim}8$\,au by assuming the companion is on a 17\,year orbit with an eccentricity close to 1 \citep[an assumption likewise made by][${\sim}$2\,au wider than the $\rm{RadVel}$ estimation of $a_{\rm{c}}{=}$6\,au in Table~\ref{tab:bestfitplanets}]{Bonsor13}. 
Further, if we assume such a body is the origin of the RV trend but was instead in a circular orbit, we can estimate its mass (in $M_{\rm{Jup}}$) as 
\begin{equation}
\label{eq:RVlim}
M_{\rm{pl}}\,\sin{i}\, > 5.6\times 10^{-3} \Big( \frac{a}{[\rm{au}]} \Big)^2 \Big( \frac{\dot v}{[\rm{m\,s^{-1}\,yr^{-1}}]} \Big)\,M_{\rm{Jup}}\,,
\end{equation}
for semi-major axis ($a$) in au, and a constant acceleration term ($\dot v$) in $\rm{m\,s^{-1}\,yr^{-1}}$. 
This estimation yields a lower limit putative RV planet mass of $0.4\,M_{\rm{Jup}}$ (at 8\,au). 
Likewise, given the lack of any significant astrometric proper motion anomaly, with Gaia DR2 and eDR3 data we can set an upper limit on the mass (for a given semi-major axis) as 
\begin{equation}
\label{eq:PMAlim}
M_{\rm{pl}} \lesssim \lvert \Delta \mu \rvert \frac{Ia^2}{G \omega \gamma(\Phi,i,\phi_v)}\,,
\end{equation}
where $\lvert \Delta \mu \rvert$ is the modulus of the difference in proper motions between the two Gaia measurement epochs (i.e., combining both RA and Dec directions), $I$ is the time interval between the two Gaia measurement epochs, $a$ is the semi-major axis, $\omega$ is the parallax, and $\gamma$ is a term that accounts for the orbital phase and viewing geometry.
Note that $\gamma$ is a term of order unity that can vary between 0.5 and 2, for which we assume the average value of ${\gamma}{=}1$.
Using the astrometric data in Table~\ref{tab:GaiaAstrometry}, this equation yields a new upper limit on the putative RV planet mass of ${\sim}120\,M_{\rm{Jup}}$, though we note that this equation is only valid for outer bodies on circular orbits and in the `long period limit', which we define and derive in Appendix~\ref{sec:AppendixB}. 
The previous Keck AO limits could only constrain this upper limit mass as ${\sim}200\,M_{\rm{Jup}}$.
In \S\ref{sec:discPlanInts}, we further this analysis by combining the planetary system constraints outlined here with our modelling of the debris disc architecture.

\begin{table}
    \centering
    \caption{Astrometric data for $\kappa\,$CrB used to derive proper motion anomaly limits (shown in Fig.~\ref{fig:planetspace}). Note that epoch 1 and 2 refer to Gaia data releases DR2 and eDR3 respectively \citep[see][]{Gaia18, Gaia21}.}
    \begin{tabular}{l|c|c|c}
         \hline
         \hline
         Parameter & Value & Error & Units \\
         \hline
         Parallax, $\omega$ & 33.23 & 0.11 & mas \\
         Proper motion (RA), epoch 1, $\mu_{\rm{RA,\,1}}$ & -8.79 & 0.18 & mas\,yr$^{-1}$ \\
         Proper motion (RA), epoch 2, $\mu_{\rm{RA,\,2}}$ & -8.67 & 0.07 & mas\,yr$^{-1}$ \\
         Proper motion (Dec), epoch 1, $\mu_{\rm{Dec,\,1}}$ & -347.76 & 0.20 & mas\,yr$^{-1}$ \\
         Proper motion (Dec), epoch 2, $\mu_{\rm{Dec,\,2}}$ & -348.35 & 0.09 & mas\,yr$^{-1}$ \\
         \hline
    \end{tabular}
    \label{tab:GaiaAstrometry}
\end{table}

\section{Discussion}
\label{sec:discussion}
Here we will piece together the observational and modelling analyses presented earlier to tie-up our understanding of the $\kappa$\,CrB planetary system as a whole, and place this in the context of other planetary systems.
Thus far, our analysis of this 2.5\,Gyr sub-giant can be summarised as follows:

a) from the millimetre imaging and modelling, there is a single, $61^\circ$ inclined, broad planetesimal belt with a position angle of ${\sim}146^\circ$ (anti-clockwise relative to North), with a peak emission radius of ${\sim}110\,$au, which with an ALMA resolution of $63.2\times 42.1$\,au, produces emission that extends over a wide radial region from the star of ${\sim}50{-}180$\,au;

b) from the scattered light imaging, there is significant dust emission that extends from a radius of at most 51\,au out to 280\,au (peaking at 73\,au) with an estimated albedo of $A{\gtrsim}10\%$, and although has a position angle on the sky consistent with that measured at millimetre wavelengths, is significantly less inclined;

c) from the RV monitoring, there is an inner planet on a 1248-day orbit, and a significant long-term RV trend, which if due to an outer body must have a mass between $0.4{-}120\,M_{\rm{Jup}}$, and an orbital radius between $8{-}66$\,au.

\subsection{Understanding $\kappa$\,CrB's planetesimal belt}
\label{sec:discussionBelt}
\subsubsection{The mass of $\kappa$\,CrB's planetesimal belt}
\label{sec:discussionBeltMass}
We start by assessing the mass of $\kappa$\,CrB's planetesimal belt based on our millimetre analysis. % in order to compare this with the population of debris discs.
We note that our models of the Band~6 emission both found a dust mass of ${\sim}0.016\,M_\oplus$ with maximum model dust grain sizes of ${\sim}2\,$cm which can be used to constrain the size of planetesimals, and thus the total belt mass. 
For example, by setting the 2.5\,Gyr system age equal to the collisional lifetime of the planetesimals in the belt \citep[see eq.3 of][]{Wyatt08}, we set a lower limit on the maximum planetesimal size of bodies in the collisional cascade of ${\gtrsim}1\,$km, and from this we can estimate a lower limit on the total belt mass of ${>}1\,M_\oplus$ \citep[assuming an $\alpha=3.5$ size distribution, as per][]{Dohnanyi69}.
The above assumes a fixed value for $Q^\star_{\rm{D}}$ of ${\sim}70\,$J\,kg$^{-1}$, a relative velocity of $660{\pm}130\,$m\,s$^{-1}$ \citep[using eq.10 of][with a fixed value of $h=0.05$]{Matra19}, a grain density of $\rho=2.7\,$g\,cm$^{-3}$ and a width 50\,au, i.e., consistent with the lower limit power-law model extent.
This total belt mass and lower-limit on planetesimal size is consistent with other debris discs at a similar age, like 1.4\,Gyr q$^1$~Eri \citep{Lovell21c}, suggesting that although this disc is indeed long-lived, it is not an outlier in having retained detectable dust over ${\gtrsim}$Gyr ages.

\begin{figure}
    \includegraphics[width=1.0\columnwidth]{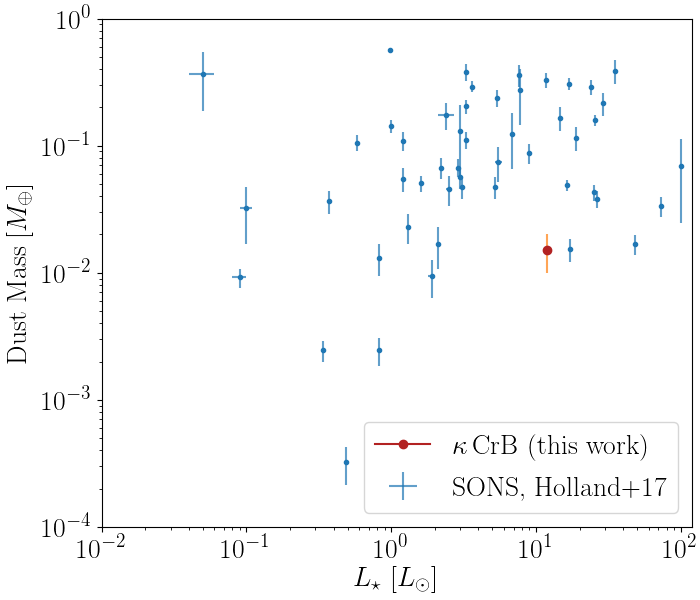}
    \includegraphics[width=1.0\columnwidth]{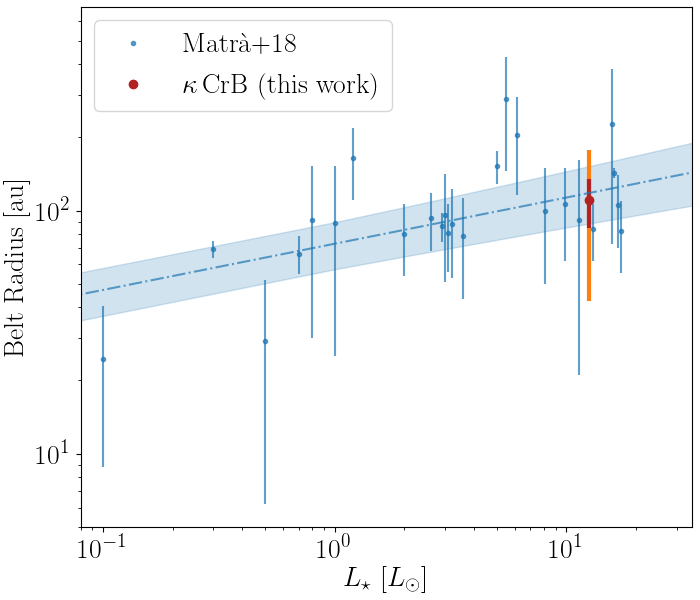}
    \caption{Top: $M_{\rm{dust}} - _\star$ for the SONS survey of debris discs \citep[showing the complete data set in blue as presented in][]{Holland17}, including the new data point for $\kappa\,$CrB (amber).
    Bottom: $R_{\rm{disc}} - L_\star$ relationship plot \citep[showing the complete data set in blue as presented in][for resolved disc radii]{Matra18}, including the new data point for $\kappa\,$CrB, along with its 50\,au $3\,\sigma$ extent (in red) and 135\,au median extent (in amber).}
    \label{fig:RdiscLstar}
\end{figure}

We next compare the modelled Band~6 dust mass of $\kappa\,$CrB with the values determined for the SONS sample of main sequence debris discs \citep{Holland17} on Fig.~\ref{fig:RdiscLstar} (top).
Whilst the SONS dust masses were determined at 850$\mu$m with a dust opacity of $\kappa_\nu{=}1.7\,$cm$^2$\,g$^{-1}$, to compare these with the dust mass derived for $\kappa\,$CrB's debris disc at 1.3\,mm it is important to use a consistent dust opacity value, thus we scaled the SONS dust masses by assuming an opacity power-law index of $\beta{=}1$ \citep[see e.g.,][which we note only alters the SONS dust masses by ${\sim}50\%$]{Hildebrand83,Beckwith90}. 
We find that the dust mass of $\kappa\,$CrB's debris disc is lower than most of the discs around less-evolved A-type stars, however this is unsurprising given that $\kappa\,$CrB is nearly an order of magnitude older than all SONS A-type stars, and has thus had longer to collisionally grind down. 
In fact, by comparing the dust mass of $\kappa$\,CrB with the mass evolution of \citet{Holland17}, i.e., the $\propto t^{-0.5}$ trend on their Fig.~34, it can be seen that the dust mass measured here is entirely in accordance with this prediction based on collisional evolution models \citep[see e.g.,][]{Wyatt07,Sibthorpe18}.

Further, by comparing the disc peak emission radius, extent and stellar luminosity of $\kappa\,$CrB with other main sequence debris discs \citep[using the data presented by][]{Matra18}, we show on Fig.~\ref{fig:RdiscLstar} (bottom) that $\kappa\,$CrB's debris belt peak radius remains strongly consistent with debris discs around main sequence A-stars.
We present on this plot both the strict and median extents of the disc (i.e., the 50\,au and 135\,au values demonstrated by the broad disc model SPL), demonstrating that both these measures of the disc extent are also consistent with main sequence A stars.

Neither of the conclusions that $\kappa\,$CrB's debris disc is consistent with the dust masses and disc radii of main sequence A-type debris discs are particularly surprising given how early $\kappa\,$CrB is into its giant branch evolution (see e.g., Fig.~\ref{fig:HRdiag} in which early stage sub-giants have comparable luminosities to main-sequence A-type stars) and that A-type stars are readily detected with debris discs across the span of their main sequence ages \citep[see e.g.,][]{Rieke05, Su06, Wyatt07}. 
Nevertheless we have shown that for the total \textit{disc} mass, total \textit{dust} mass and disc radius, that $\kappa\,$CrB's disc remains consistent with the population of less-evolved debris discs around A-type main sequence stars in the context of collisional evolution, despite its older ${\sim}2.5\,$Gyr age.
Future observations of sub-giants/giant branch stars to search for the presence of debris disc dust may uncover whether or not the standard collisional evolution models remain applicable around stars that have evolved similarly to, or beyond the stage of $\kappa\,$CrB.
We further note that the levels of dust present are significantly higher than those in the Solar System's Kuiper Belt \citep[see, e.g.,][]{Wyatt08, Booth09}.
From this we highlight the same point made in \citet{Bonsor13} that this strongly implies there very likely has been no `Late Heavy Bombardment' type event in the planetary system of $\kappa\,$CrB.

\subsubsection{The inclination of $\kappa$\,CrB's planetesimal belt}
\label{sec:discussionBeltInc}
Unusual about the observations of $\kappa\,$CrB is the inclination discrepancy between i) the millimetre Band~6 ALMA and far-infrared \textit{Herschel} images (which both suggest this to be ${\sim}60{-}61^\circ$), and ii) the scattered light image, which is better fitted with models inclined at ${\sim}30{-}40^\circ$.
%We earlier ruled this out as being due to a scale height - inclination dependency (see $\S$\ref{sec:modelHST}). 
Assuming that there are no underlying issues with either data set, here we explore three hypotheses that may explain this inclination discrepancy.

\textit{Radial scattering anisotropy dependence?}
One possibility to explain the inclination discrepancy is that the scattering factor (that determines the level of dust emission anisotropy) has a radial dependence.
This could have the effect that emission at increasingly greater distances from the star becomes preferentially scattered, which consequently reduces the assessment of the scattered light dust inclination when modelled with a fixed scattering factor (as was conducted in $\S$\ref{sec:modelHST}).
This effect is likely to be most pronounced in broad discs where there are a wide range of radial distances over which differences in scattering factors must be accounted for.
Indeed based on the best-fit models using single-valued Henyey-Greenstein functions presented in \S\ref{sec:modelHST}, to model $\kappa\,$CrB's scattered light emission with a \textit{single} phase function, such radially-dependent scattering factors are likely required (given the large residuals that remain after model subtraction).
The problems associated with modelling scattered light observations of debris discs with single-valued Henyey-Greenstein parametrisations are well known \citep[see e.g.,][]{Stark14, Milli17}. 
Whilst in addition to the HG models, we inspected models based on Mie-theory, it may still be the case that the inclination discrepancy suggests a better model is required, for example, with a radially varying phase scattering function.
We note that confirming whether or not this can fully account for the measured inclination discrepancy requires additional modelling work beyond the scope of this study with more complexity than the simple scattered light models presented here. 

\textit{A dust-size dependent scale height?}
Another possibility is that $\kappa\,$CrB's micron-sized dust is being vertically `puffed up' more than the dust observed at millimetre/far-infrared wavelengths, which could have the effect of lowering the estimated inclination at sub-micron wavelengths.
Scale heights in debris discs have a number of origins, such as being from their primordial planetesimal distribution or from planetesimal scattering by large embedded bodies \citep[for a discussion on this in the case of the edge-on disc, AU~Mic, see][]{Daley19}.
The combination of radiation pressure and collisional processes can additionally induce larger scale heights in smaller grains than larger ones. 
For example, since smaller grains are blown on to eccentric orbits by radiation pressure, these have higher impact velocities relative to larger grains, which due to momentum conservation result in these receiving larger impact `kicks' during collisions with larger grains/planetesimals.
The result of this was modelled by \citet{Thebault09}, in which small grains close to the blowout-size were shown to be more vertically puffed up than the largest grains, i.e., to average inclinations of ${\sim}8\degr$, versus just ${\sim}0.5\degr$ for the smallest and largest test particles respectively.
Therefore whilst this effect may not fully account for the ${\sim}20{-}30\degr$ difference in the modelled scattered light inclination and that observed in the far-infrared/millimetre, this may contribute to the discrepancy.
We therefore recommend further collisional modelling of the $\kappa\,$CrB system to constrain the extent to which collisions and radiation pressure can induce dust-size dependent scale heights in this debris disc. 

We note in addition that whilst radiation pressure may not be responsible for the inclination discrepancy, radiation pressure does naturally explain how the observed scattered light halo extends out to ${\sim}280$\,au, whereas the millimetre dust is not observed significantly beyond ${\sim}$180\,au.
In other words, whilst radiation pressure may not dominate the vertical distribution of the micron-sized dust, it does have a significant effect on the radial structure.

\textit{Misaligned precessing belts?}
The previous two hypotheses share the feature that their dust belts originate from the same distribution.
Here we suggest the possibility that these discs are instead part of different belt distributions, i.e., that the millimetre/far-infrared and sub-micron belts have physically different inclinations, with the parent belt (traced in the millimetre) precessing at a rate different to that of the smaller dust in scattered light halo \citep[in a mechanism similar to that discussed in][]{SendeLohne19}. 
Whilst this might be at odds with the perceived alignment in the position angles of the dust observed with ALMA, Herschel and HST (which we find are all within the range of 142--152$\degr$), for example, since any precession would be likely to misalign these belts in \textit{both} inclination and position angle, this could be a chance encounter where these distributions are aligned in position angle, but not in inclination. 
On the other hand, higher resolution imaging and more detailed modelling may indeed find that the position angles are less well aligned than our existing modelling procedure determined.
Ultimately, there remain uncertainties with all three of these scenarios, and we briefly outline further work necessary to understand the nature of this inclination discrepancy.

In our analysis of the scattered light emission, the greatest uncertainty that remains is the location of the inner edge of the dust. 
Since this lies within the HST imaging artefacts, the scattered light dust inclination derived in this work is model dependent: a direct measurement of the belt inner edge with scattered light imaging could however determine this uniquely.
We therefore suggest new scattered light observations of $\kappa\,$CrB, e.g., with \textit{SPHERE} or \textit{GPI}, to resolve and directly measure this inner edge, derive more accurately the scattered light inclination, and re-evaluate the hypotheses outlined here.

\subsubsection{Asymmetries in $\kappa\,$CrB's debris disc?}
\label{sec:discussionClump}
In our modelling section, we showed that the millimetre emission residual maps (Fig.~\ref{fig:kCrB_modsResids}) all contained the same $30{-}40\,\mu$Jy clump to the south of the belt (at a relative RA, Dec location of -1.0, -5.0).
Perhaps simply, given all of the residual maps likewise contain \textit{negative} residual emission at the same level (e.g., see RA, Dec -5.0, -2.5) we cannot rule this out as being due to noise.
On the other hand, this emission could be real, and we explore three hypotheses, namely that this is emission from a background millimetre galaxy (MMG) or dust, either from a planetesimal collision or in a circumplanetary orbit, and suggest future work that could distinguish between these scenarios.

\textit{A background galaxy?}
We firstly assess the more likely of the three options, that this is emission from an MMG.
For example, based on the galaxy counts of \citet{Simpson15}, we find the probability that an MMG with ${\geq}0.035$\,$\mu$Jy falls within a $60^\circ$ inclined ellipse with a semimajor axis of 5.0'' exceeds order unity, in other words, it is not surprising that at the depth of our observations a significant clump of emission is coincident with the debris disc.
Such a hypothesis can be tested since $\kappa\,$CrB is a high-proper motion star, whereas the location of an MMG is fixed on the sky.
However, given the proper motion of $\kappa\,$CrB is $\mu_{\rm{RA,Dec}}{=}(-8.79{\pm}0.18,-347.77{\pm}0.20)$\,$\rm{mas}\,\rm{yr}^{-1}$ \citep[see][]{Gaia18} and the resolution of our imaging (${\sim}2.1\times1.4''$), this hypothesis can only be tested at the 3$\,\sigma$ level 18 years on from these 2019 observations, in December 2037. 

\textit{A planetesimal collision?}
Secondly, we explore the interpretation that this clump has a planetesimal origin and estimate the size of the body required to produce this, by first inferring the dust mass required to produce 35$\mu$Jy emission as ${\sim}4\times10^{-4}\,M_\oplus$, i.e., using the same ratio of the total disc flux to the total dust mass.
To arise from the complete destruction of a single planetesimal, such a body would need a diameter D${\gtrsim}1200\,$km, assuming 1) a grain density of $\rho=2.7\,$g\,cm$^{-3}$ and 2) that the planetesimal breaks up only into the millimetre-centimetre-sized grains that show up as a clump in the emission.
Although this is consistent with the lower-limit maximum planetesimal size estimated earlier, this is a factor ${\sim}1000\times$ larger.
We note that this planetesimal size is a lower limit since the catastrophic break up of a planetesimal is likely to produce a cascade of different sized particles, whereas we have only considered the millimetre-centimetre sized grains. 
Without repeating the analyses of \citet{Krivov21} and \citet{Lovell21c}, we simply re-state here that if such D${\gtrsim}1200\,$km-sized bodies are part of a collisional cascade with a \citet{Dohnanyi69} size distribution, this would suggest a \textit{total} debris disc mass that is substantially larger than the ${\sim}1\,M_\oplus$ estimated earlier, for example, with a mass of many tens of $M_\oplus$, based on $M_{\rm{disc}} \propto D^{1/2}_{\rm{max}}$.
Since a more realistic collision at this location would produce significant levels of micron-sized dust as part of a cascade which would be blown out \citep[a possibility discussed by][]{Grigorieva07}, this could provide a simultaneous explanation to the inclination discrepancy discussed in \S\ref{sec:discussionBeltInc}. For example, whilst the millimetre-sized grains would spread around the orbit of the parent planetesimal, the micron-sized grains would be preferentially blown out, and thus cause the observed northeast-southwest asymmetry seen in the scattered light emission.

\begin{figure}
    \includegraphics[width=1.0\columnwidth]{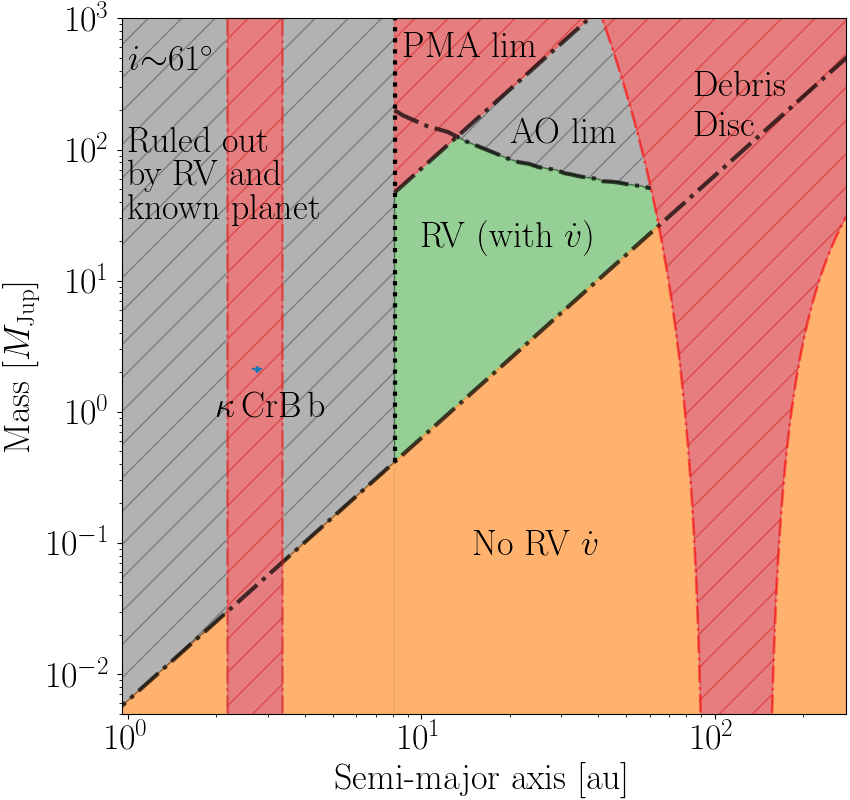}
    \caption{Shown here are the possible places that planets can/cannot reside (unhatched/hatched regions respectively) within the $\kappa\,$CrB system, based on all the constraints set out in this work, in mass-semimajor axis space (for an assumed inclination of $61^\circ$) and the location of the known planet, $\kappa\,$CrB\,b. 
    In green is the area that a companion \textit{must} exist in order to produce the observed RV trend, and in amber is the area that planets can exist without affecting any observations.
    In grey are the areas that planets cannot exist, as ruled out by \citet{Bonsor13}, i.e., due to both RV and AO limits.
    In red are the areas that planets cannot exist based on this work, ruled out by dynamical arguments, i.e., the chaotic zones of $\kappa\,$CrB\,b, and putative planets on the inner and outer edges of the debris disc. }
    \label{fig:planetspace}
\end{figure}

\textit{Massive circumplanetary rings?}
Finally we consider the possibility that this is unresolved circumplanetary dust orbiting an outer body with the same dust mass as inferred from the planetesimal origin hypothesis, i.e., ${\sim}4\times10^{-4}\,M_\oplus$.
This level of mass is just ${\sim}1/30$ that present in the moon, or $100{-}200\times$ that present in Saturn's rings, and consequently, would not require a particularly massive body to retain this level of dust.
Our modelled limits on the size of the millimetre debris disc can provide additional constraints on the mass of such a body based on the outer edge debris disc dynamical limits, as plotted on Fig.~\ref{fig:planetspace} and which we discuss in detail in \S\ref{sec:discPlanInts}.
For now we note that the limits set by the debris disc model shows that the mass of any planet \textit{external} to the planetesimal belt must be ${\lesssim}3\,M_{\rm{Jup}}$ at the 205\,au distance of the clump to the star (assuming this orbits in the same plane as the belt). 
In other words, whilst it may not be necessary to explain the outer edge of this debris disc via planet sculpting, if this clump is due to circumplanetary dust around a massive outer planet (which we emphasise would be a different body to the one causing the RV trend) then it may additionally constrain the disc outer edge location.
Better constraints on the disc outer edge could therefore more tightly constrain the outer planetary system region. 

Since the planetesimal collisions and circumplanetary dust hypotheses have a common origin in the emission being due to dust these are difficult to distinguish, even with long-term, high-resolution millimetre monitoring.
Although dust formed in a collision would gradually spread around its orbit over time whereas dust in a circumplanetary ring system would remain coincident with its host-planet, we should expect the shape of the the millimetre emission of these two scenarios to diverge over time. However, at ${\sim}205$\,au orbital timescales are very long and it may take many hundreds to thousands of complete orbits for dust formed in a planetesimal collision to spread sufficiently to be measurable \citep[see][for a discussion on collisional spreading timescales]{Jackson14}.
Therefore, since it could take many thousands of years to accurately distinguish between these origins in the millimetre, detailed analysis of this emission in other wavebands is likely to be necessary.

\subsection{Planet-disc interactions in $\kappa$\,CrB's planetary system}
\label{sec:discPlanInts}
Having discussed the planetesimal belt in isolation in $\S$\ref{sec:discussionBelt}, here we extend our analysis of the wider system by considering the impact that the plausible planet architecture has on the planetesimal belt structure by considering possible planet-disc interactions \citep[as studied extensively over recent years for the case of axisymmetric and asymmetric discs, see e.g.,][]{Wyatt05,Mustill09,Shannon16,LynchLovell21}.

We start by introducing Fig.~\ref{fig:planetspace} which shows the parameter space within which \textit{any} planets/companions in this system can reside, assuming the RV trend is caused by a massive outer companion. We define 3 categories, namely where an outer body \textit{must} reside (in green), \textit{can} reside (in amber), and \textit{cannot} reside (in grey or red, depending on these constraints being based on \citet{Bonsor13}, or this work, respectively). 
The grey regions are set by i) the previous Keck II adaptive optics limits, and ii) the RV data (using the 2004-2021 baseline) which strictly constrains the types of planet that can reside within 8\,au (see \S\ref{sec:planetArch}).
The amber region is set by Equation~\ref{eq:RVlim}, above (below) which planets would (would not) induce an RV acceleration trend.
The red limits are set by three dynamical constraints. 
Firstly, that other planets cannot reside within the orbital instability region set by $\kappa\,$CrB\,b (i.e., within its chaotic zone, where we note the inner and external chaotic zones of a planet as $\Delta a_{\rm{cz}} {=} {\pm} 1.5 a_{\rm{pl}} (M_{\rm{pl}}/M_\star)^{0.28}$ \citep[see e.g.][]{Quillen06}, where $a_{\rm{pl}}$ is the planet semi-major, and $M_{\rm{pl}}$ is the planet mass).
Secondly, that planets cannot be present in the parameter space within which these would cause an observable proper motion anomaly between the Gaia DR2 and eDR3 epochs (`PMA lim').
Thirdly, that planets cannot be present where these would have chaotic zones that would intersect with the inner and outer edges of the debris disc.
The green region is then bounded by all such conditions, and defines the parameters of a body that would cause the RV trend.
This has two important implications.
Firstly, that if the RV trend is due to an outer body, this must be massive, with $0.4{-}120\,M_{\rm{Jup}}$, and lie between $8{-}66\,$au, e.g, this could be consistent with a range of bodies, such as an eccentric exo-Saturn or a brown dwarf at tens of au.
Secondly, that additional planets \textit{can} be present without inducing an RV acceleration trend in the RV data if these reside below the RV trend limit, i.e., in the amber regions, meaning that inner terrestrial planets, and even giant planets at tens of au can stably reside in this system without being detected.
Nevertheless, we note that if the RV trend is not due to an outer body, the constraints on where planets may reside differ from Fig.~\ref{fig:planetspace}. 
For example, although the red and grey AO limit regions would remain, any regions bounded by the RV data would be unconstrained.

\subsubsection{Dynamical constraints: more than 2 planets?}
Since scattered light sub-micron emission does not trace the parent planetesimal belt as accurately as millimetre emission, the constraints we place on the planetary architecture are derived from the Band~6 ALMA data.
The millimetre modelling in $\S$\ref{sec:modelling} showed that whilst 1.3\,mm emission begins at ${\sim}50\,$au, this is consistent with a broad belt between 93-151\,au (though the outer edge is a lower limit, e.g., see discussion of model SPL in $\S$\ref{sec:modellingresults}).
The presence and morphology of this disc provides constraints on the planetary architecture in three ways. 
Firstly, the presence of a large cavity between the star and the disc inner edge suggests that this was caused by planet-sculpting from unseen inner planets.
Secondly, since the chaotic zones of planets cannot overlap with the location of the planetesimal belt, there is likely a dearth of planets (above a given mass) in this region.
Thirdly for the disc to be producing visible levels of dust, it is possible that this disc is being stirred by a planet.
There is however only one known planet, the inner exo-Jupiter $\kappa\,$CrB~b, and this planet alone could not be responsible for the observed disc architecture.
For example, we rule out the possibility that belt stirring can be achieved by this inner exo-Jupiter since the secular stirring timescale \citep[see eq.15 of][]{Mustill09} is ${\gtrsim}1000\times$ the age of this system. %, i.e., it is implausible that a planet with low eccentricity at 2.8\,au is stirring a belt at ${\sim}110\,$au.
The combination of these three conditions (in addition to the known RV acceleration trend, see $\S$\ref{sec:planetArch}) thus argues in favour of the presence of additional outer planets, which we discuss further below.

We first consider a scenario in which planets underwent no significant migration, and note first that the limits set by the debris disc stability on Fig.~\ref{fig:planetspace} are based on the requirement that planetary chaotic zones \citep[i.e.,  $\Delta a_{\rm{cz}} \pm 1.5a_{\rm{pl}}(M_{\rm{pl}}/M_\star)^{0.28}$, see][]{Quillen06} cannot overlap with the planetesimal belt which spans from 93-151\,au.
The disc inner edge thus sets two constraints on the type of companion/s that may be responsible for causing the observed cavity via i) sculpting the belt inner edge at 93\,au and ii) depleting planetesimals from a few au outwards to 93\,au.
Assuming that planetesimals originally formed at all distances from the star, in a scenario where no planets migrated, we find that in order to explain the RV trend, the extent of the cavity and the disc inner edge location, multiple additional planets within the inner cavity are required.
For example, the widest orbiting companion within the cavity that could carve the inner edge (i.e., a massive body at 66\,au, see Fig.~\ref{fig:planetspace}) can only clear planetesimals from 46\,au out to 93\,au, i.e., such a companion cannot clear planetesimals sufficiently far inwards and thus additional planets would be required to do so in the inner tens of au.
Indeed, there are no scenarios in which a single companion in addition to $\kappa\,$CrB~b can carve the morphology of the debris disc (without requiring any planetary migration, which we discuss further in \S \ref{sec:discplanmig}).

To constrain the number and mass of any such planets at tens of au responsible for carving the cavity, we consider two limiting scenarios based on the putative RV companion residing at either end of its constrained location.
If at 8\,au or 66\,au the putative RV companion would have chaotic zones either between 5-13\,au or 46-93\,au respectively. 
By using equations 4 and 5 of \citet{Shannon16} and the system age of 2.5\,Gyr, we can estimate possible planetary architectures by assuming that these planets would be responsible for the disc cavity \textit{either} between 5-46\,au (i.e., in the case that there is a massive companion at 66\,au) or between 13-93\,au (i.e., in the case that there is a massive companion at 8\,au). 
Respectively these were found to either require 13 planets each with a mass ${>}$0.7\,$M_\oplus$ or 8 planets each with a mass ${>}$2\,$M_\oplus$.
In either of these two scenarios, all of these bodies would be in the amber region of Fig.~\ref{fig:planetspace}, and the putative RV companion in the green region.
We note here however, that since the cavity may have formed much earlier in the system's lifetime (and the equations of \citet{Shannon16} are based on the age of the system) fewer planets with a greater total mass could have been responsible for the observed cavity i.e., the above planet mass estimates should be seen as lower limits.
Since the putative RV companion could however be anywhere between 8-66\,au (rather than at either limit) these two example calculations serve to demonstrate that in addition to $\kappa$\,CrB~b and the putative RV planet, a significant number of additional planets may be required to produce the observed disc morphology if planet sculpting from non-migrating planets is the main cause of the observed cavity.

\subsubsection{Planet migration: just 2 planets?}
\label{sec:discplanmig}
An alternative hypothesis is that this system formed with two planets; the inner eccentric exo-Jupiter, $\kappa$\,CrB~b, and the outer RV companion which underwent significant migration.
For example, this outer body could have formed in the inner regions of this system, and migrated outwards to 66\,au following, e.g., a planet-planet interaction with $\kappa$\,CrB~b, or another dynamical instability, clearing the full cavity and carving the 93\,au inner belt edge.
Indeed it has been shown that planet migration over many 10s of au is possible in systems with $N{\geq}$2 planets in which one of these migrates out through a disc \citep[see e.g.,][]{Martin07}.
As such, a body migrating through the belt would interact significantly with the belt planetesimals, either accreting, ejecting or trapping these in resonances. 
Potentially problematic for such a scenario is that planetesimals trapped in resonances could induce significant sub-structure in the debris disc \citep[e.g., such as clumps, see][depending on parameters such as the migration rate and planet mass]{Wyatt03}, which we do not observe.
Although at the depth of our millimetre observations we see no evidence for any significant asymmetries in the emission co-located with the belt that might indicate early planet migration, future deep observations may uncover sub-structure related to such a dynamical history.
On the other hand, if such a migration happened far earlier in $\kappa\,$CrB's lifetime, the population of planetesimals in the resonant zones may have depleted sufficiently that such asymmetric imprints no longer remain observable, i.e., since such observable asymmetries can be smoothed over.
For example, in systems where planetesimals start with moderate eccentricities, asymmetries become less pronounced over time \citep[see e.g.,][]{Reche08}. 
Thus, deeper images of $\kappa\,$CrB's debris discs may still be insufficient to exclude this type of migration history.

\subsubsection{What will become of this debris disc?}
The fate of $\kappa\,$CrB's debris disc as the star ascends the giant branch is intimately bound with the location of its planets.
\citet{Bonsor10} showed that as a result of post main sequence stellar evolution and the adiabatic stellar mass loss of A-type stars at the end of the giant branch stage, the enlargement of planet and planetesimal orbits (which can expand by the same fraction of stellar mass lost) has implications for the depletion of debris discs.
For example, the fractional increases in the orbital distances are of the order 2-3 towards A-type stars, thus further increasing the size of the chaotic zones of planets and the opportunity for resonance-overlap instabilities, but also the radii of the planetesimal orbits.
This means that if a massive companion is responsible for the sculpting of the inner edge of this debris disc, the disc may be significantly disrupted by such a body as the star loses mass towards the end of the giant branch stage.
Such disruptions present significant opportunities for the \textit{inward} scattering of planetesimals and thus their break-up/sublimation near the stellar surface, thus the $\kappa\,$CrB system could very well be one of the precursors to the population of bright `polluted' white dwarfs seen at much later evolutionary epochs \citep[see][]{Zuckerman03,Zuckerman10,Koester14,Veras14}.
On the other hand, the presence of the inner planet $\kappa\,$CrB~b (which at such an evolved epoch may instead reside at ${\sim}$6-8\,au) could efficiently eject such inwardly scattered planetesimals, and thus the flux of planetesimals on star-grazing orbits may be consequently very small.
New modelling based on the constrained planetary system architecture of this work could therefore explore the level of pollution expected at the white dwarf stage.

\subsubsection{How does $\kappa$\,CrB's planetary architecture compare to others?}
Planet formation in the $\kappa$\,CrB system must have been reasonably efficient. 
Even just considering the total planet mass of the confirmed exo-Jupiter $\kappa$\,CrB~b, ${>}2{\times}$ the amount of material formed planets in the young protoplanetary disc of $\kappa$\,CrB than did so in the Solar System (though we note, $\kappa$\,CrB may be nearly twice as massive as the Sun).
This however could be as high as ${\sim}120{\times}$ that of the Solar System, e.g., if the RV acceleration trend is due to a planet at the upper mass limit of Fig.~\ref{fig:planetspace}. 
Whilst compared with our G-type Sun this suggests that planet formation may have been substantially more efficient around $\kappa$\,CrB, this may instead reflect the larger average mass of material available to form planets within protoplanetary discs that is available around intermediate mass stars \citep[see, e.g., ][]{Ansdell16, Pascucci16}. 
Alternatively, planet formation efficiency around $\kappa$\,CrB may also be substantially lower than other intermediate mass stars.
For example, around the 30\,Myr A-type star, HR\,8799, at least four giant planets formed with a total mass of ${\sim}$30\,$M_{\rm{Jup}}$ \citep{Marois10}.
If the putative RV companion is towards the lower end of the constrained mass bounds of Fig.~\ref{fig:planetspace}, then planet formation may have been over an order of magnitude more efficient around HR\,8799 than $\kappa$\,CrB.
Nevertheless, with only loose constraints on the total planet mass in the $\kappa$\,CrB system, there remains a large uncertainty on whether the planet formation efficiency is indeed discrepant with known planetary systems around stars that could have formed with comparable levels of primordial material.

It is further instructive to compare the location of planets around $\kappa$\,CrB with other systems.
We firstly do so with reference to the snowline.
For the simple assumption that dust within planetary systems behaves as a blackbody, we can estimate the location of the snowline from $r_{\rm{snowline}}=(278.3/T_{\rm{snow}})^2L^{0.5}_\star$. 
For a sublimation temperature of $T_{\rm{snow}}{=}170$\,K and $\kappa$\,CrB's luminosity $L_\star=12\,L_\odot$, this equation would place the snowline at 9.3\,au, though we note that earlier in the system's history when planets were forming this will have been closer to 6\,au.
This would mean that the known planet $\kappa$\,CrB~b is inside of this, with the putative outer RV companion outside of this. 
This means that irrespective of the \textit{mass} of planets formed around  $\kappa$\,CrB, the broad architecture of this is different to that of HR\,8799 in which all four giant planets orbit with radii significantly beyond their snow lines.
Although this may indicate that $\kappa$\,CrB and HR\,8799 have distinct planet formation histories, the observed architectural differences may also be due to the young age of the HR\,8799 system in which further orbital architecture evolution will occur, for example, via planet migration (e.g., through planet-planetesimal belt interactions) or via planet-planet scattering events.
In contrast, whilst the mass levels of the bodies in the Solar System and $\kappa$\,CrB are at least a factor of 2 different, the location and type of bodies may be entirely consistent.
For example, Fig.~\ref{fig:planetspace} demonstrates that planets consistent with an exo-Earth (lower left amber region), an exo-Jupiter ($\kappa$\,CrB\,b), an exo-Saturn (the RV trend companion at the lower left of the green region) and multiple exo-Uranus/exo-Neptune analogs (in the middle amber region) are all consistent with the RV data, and need only a few more Earth-mass planets tens of au beyond those to carve the disc cavity and inner edge.
In other words, the planetary architecture of $\kappa$\,CrB may be a slightly broader and more massive version of the Solar System, albeit with a significantly more massive planetesimal belt.
As such, with tighter constraints on the planetary architecture of $\kappa$\,CrB, Solar System formation models may provide significant insight into how this planetary system formed and evolved.

\subsection{CO gas: constraints on giant branch sublimation?}
\label{sec:discussionCO}
\begin{figure}
    \includegraphics[width=1.0\columnwidth]{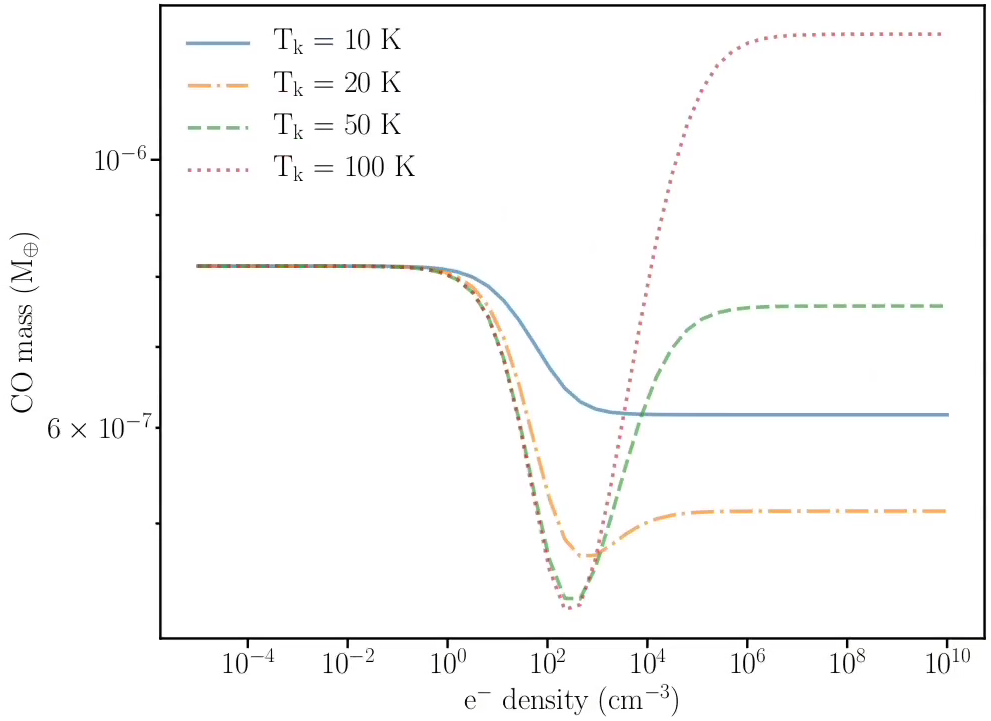}
    \caption{Interpreted limits on the CO gas mass (based on the upper-limit CO flux) plotted as a function of the unknown density of electron collision partners, for four assumed temperatures (10, 20, 50 and 100\,K, in blue/solid, amber/dash-dotted, green/dashed, and purple/dotted respectively.}
    \label{fig:COmass}
\end{figure}

Having fully explored the planetary system dust observations and RV planet modelling, we attempt to constrain the planetary system composition.
In $\S$\ref{sec:COflux}, by analysing the emission near the J=2-1 transition line frequency, we demonstrated that if any CO gas is present in the $\kappa\,$CrB planetesimal belt (or closer to the star), it must be below our detection limits. 
However, the upper limit flux ($25.4\,\rm{mJy\,kms^{-1}}$) can be used to determine an upper limit on the total mass of CO gas present based on the gas excitation conditions \citep[i.e., as set by the kinetic temperature, density of electrons, and the radiation environment, see][for further details]{Matra15}. 
For the CO J=2 rotational level and based on the system being in non-local thermodynamic equilibrium (NLTE), we used the code of \citet{Matra15, Matra18} including fluorescence to estimate the CO gas mass at four temperatures (i.e., 10, 20, 50 and 100\,K, consistent with the expected temperatures of outer, cool debris discs) for the disc with a peak emission radius at ${\sim}110\,$au, and the star with a stellar flux consistent with the SED (i.e., see $\S$\ref{sec:SED}).
Fig.~\ref{fig:COmass} shows the outcome of these calculations for each of the four temperatures over a range of electron densities.
From this plot, we find an upper limit on the CO gas mass between $(4.2{-}13)\times10^{-7}\,M_\oplus$ (i.e., based on the 100\,K minima and maxima, which correspond to moderate and high electron collision partner densities). 
We note that electron densities range from low values (i.e. ${\sim}10^{-4}\,$cm$^{-3}$) where excitation is radiation-dominated, through to high values (i.e. ${\sim}10^{10}\,$cm$^{-3}$) where excitation is collision-dominated. 

From this we can estimate how much CO may be present in the solid planetesimals via two scenarios: i) that all CO is released via collisions, or ii) that all CO is released via sublimation.
Considering the first of these, we apply equation 2 of \citet{Matra17b} to bound the $\rm{CO}$ and $\rm{CO_{2}}$ ice-fraction of planetesimals.
We estimate this limit between $f_{\rm{CO+CO_{2}}}{<}85{-}95\%$, based on the 110\,au belt radius, lower-limit on the disc extent of 50\,au, fractional and stellar luminosities consistent with $\S$\ref{sec:SED}, a stellar mass of $1.8\,M_\odot$, a CO photodissociation timescale of 120\,yr \citep{Visser09}, and the range of CO gas masses determined above.
We note that although this range of ice fractions sets only loose constraints on the composition of planetesimals, it is strongly dependent on both the belt extent and photodissociation timescale, for which a belt with a narrower extent and well-shielded CO (and thus longer CO photodissociation timescales) would reduce $f_{\rm{CO+CO_{2}}}$ substantially. 
Nevertheless, the range of ice fractions determined here remains consistent with Solar System comets, readily found in the range of $f_{\rm{CO+CO_{2}}}{\sim}5{-}20\%$ \citep{LeRoy15}. 

An alternative scenario could be that any CO that is present in the $\kappa\,$CrB debris disc is released via the sublimation of CO-ice, as the star ascends the giant branch and planetesimal and grain surfaces are warmed up.
However, based on equation 15 of \citet{Jura04} we find at a temperature of 70\,K that the total area of dust in the $\kappa\,$CrB debris disc is a factor of $4{\times}10^{10}$ too low to liberate detectable levels of CO, i.e., consistent with our non-detection.
In estimating this value, we assume a number of parameters.
Firstly, that the belt has a temperature of 70\,K based on the blackbody temperature of dust at the peak disc radius of 110\,au around a $12\,L_{\odot}$ star.
Secondly, that the total cross sectional area of dust present \citep[see equation 4 of][]{Wyatt08} is ${\sim}7.6$\,au$^2$, for which we use the millimetre disc radius, $r{=}110\,$au, and disc fractional luminosity, $f{=}5{\times}10^{-5}$.
Thirdly, that CO survives over a 120\,yr photodissociation timescale such that there is no shielding of CO \citep[see e.g.,][]{Visser09}.
Fourthly, that the mass production rate is consistent with Kuiper Belt Object sublimation model predictions, i.e., with a mass loss rate per unit surface area as a function of temperature, $\dot \sigma_0{=}3.8{\times} 10^{8}$\,g\,cm$^{-2}$\,s$^{-1}$\,K$^{-1/2}$, \citep[see][]{Jura04}.
Finally, that to be observable, a CO gas mass of ${\sim}1{\times}10^{-6}\,M_{\oplus}$ would be required with a CO-ice fraction of 10\%, and that all liberated CO begins in an ice-phase.
We therefore note that given the very low levels of CO that can be liberated via this sublimation mechanism around $\kappa\,$CrB, any CO that is present is likely to be dominated by that produced via collisions, as is the case for main sequence debris discs.

However, the CO-ice sublimation temperature dependence is sharp, e.g., temperatures beyond 100\,K can liberate CO at a much faster rate than is possible at 70\,K.
The evolutionary timescale for ascending the giant branch can take tens to hundreds of Myr, and so a period of sublimation-dominated CO gas production around $\kappa\,$CrB has not yet been reached, however more luminous stars may have already reached this stage.
We note for the average debris disc radii of AFGK type stars this condition is only met when $L_\star$ is in the hundreds of $L_\odot$, and so is not possible at the sub-giant stage.
Given the relationship between temperature and sublimation rate, and between the reduction in planetesimal size with temperature \citep[see equations 15 and 21 of][respectively]{Jura04}, such high disc temperatures increase the contribution to the total CO available via sublimation from \textit{larger} planetesimals.
Whereas at temperatures ${\lesssim}100$\,K, the total CO mass liberated via sublimation is dominated by the contribution from ${\sim}$micron-sized grains.
As such, on the giant branch, there may arise a transition in the dominant mechanism by which CO is released, from collisions to sublimation, and towards a CO production rate sufficiently high that CO becomes readily observable.
It may thus be the case that whilst CO could be rare to detect at the sub-giant stage, studies of giant stars detect CO gas in abundance.
We note that very few observations of debris discs around giant stars have been made, and thus disc evolution during this stage should be probed further.
We therefore suggest new observations and further detailed modelling work to better constrain CO production during giant branch ascent, accounting for the CO release contribution from both collisions and sublimation, the effect of CO photodissocation, and for debris discs with varying radii and widths.

\section{Conclusions}
\label{sec:conclusions}
New images and analysis of the sub-Giant star $\kappa\,$CrB have been presented at 1.3\,mm with ALMA and at 0.6$\,\mu$m with HST, which demonstrate the existence of a debris disc with peak emission radius of $110\,$au with an extent between (at least) 50-180\,au in the millimetre, and between (at least) 51-280\,au in scattered light.
Given $\kappa\,$CrB has now evolved off the main sequence, this is therefore the most evolved debris disc resolved with ALMA to date, and demonstrates the feasibility of imaging post-main sequence debris discs.

By modelling the Band~6 ALMA data, we have shown that the continuum emission is consistent with being due to a single broad planetesimal belt with an extent ${>}50\,$au and inner edge starting at 93\,au.
We show that the debris disc of $\kappa\,$CrB is consistent with the population of discs around main sequence A-type stars (which this star evolved from) and has a mass consistent with expectations from collisional evolution models, i.e., sub-giant branch evolution may not yet have significantly influenced its debris disc properties.

We presented an updated analysis of the stellar RV over a baseline of 17\,yrs and re-confirmed the presence of an inner eccentric exo-Jupiter at ${\sim}$2.8\,au, and an acceleration trend in the residual RV data.
If the RV acceleration trend is due to an outer body, we found that it must have a semi-major axis between $8{-}66$\,au and a mass between $0.4{-}120\,M_{\rm{Jup}}$, i.e., this would need to be at least a wide-orbiting giant planet. 
As such, combined with the known exo-Jupiter, this suggests that planet formation towards $\kappa\,$CrB was efficient, forming ${\geq}2$ giant planets.

Further, we used the constraints set by the RV data and the debris disc in the millimetre to consider the architecture of the complete planetary system, showing that the cavity between the star and disc can be explained by a single massive outer companion and a string of lower mass bodies.
This analysis demonstrates the unique constraints disc observations can place on planets. 

We extended our single broad belt model to investigate the HST scattered light emission of $\kappa\,$CrB and found this is best-fitted with highly anisotropic dust (with Henyey-Greenstein factor $g{=}0.8{-}0.9$) but with a significantly lower inclination (i.e., 30--40$\degr$) than the millimetre (i.e., ${\sim}60\degr$), despite the HST and ALMA images showing broadly consistent position angles, and emission scale sizes.
We explored three possibilities; that our scattered light model is insufficiently detailed, that the sub-micron dust has a much greater vertical scale height than the millimetre dust, or that given the old age of this system, these are due to two misaligned precessing distributions.

Since there is a clump in the south of the disc in the millimetre images, we explored a number of scenarios as to the origin of this.
Although it seems most plausible that this is either noise, or a background millimetre galaxy, we discussed the potential that this is also a clump of dust, and by constraining its mass, discuss the implications of this being either due to a planetesimal collision in the belt, or dust in a circumplanetary ring system.

By further analysing emission near the CO J=2-1 spectral line, we also report no evidence of CO gas.
With our uncertainties, we constrain the total CO gas mass as $M_{\rm{CO}}{<}(4.2{-}13) \times10^{-7}\,M_\oplus$.
We find this is consistent with levels expected from planetesimal collisions (for planetesimals with Solar System comet-like compositions), or those associated with the sublimation of pure CO-ice as $\kappa\,$CrB begins to ascend the giant branch.

Deeper observations in the future will allow tighter constraints to be placed on the morphology of the belt at sub-micron and millimetre wavelengths, search deeper for any gas that may be present as this system begins to warm up as the star evolves off the main sequence, and tighten the constraints placed here on the wider planetary system.

\section*{Acknowledgements}
We thank the anonymous referee for their insightful comments that greatly improved the clarity of this manuscript.
We thank Luca Matr\`a for providing code with which CO gas mass limit estimates were derived.
We gratefully acknowledge the support and expertise provided by ALMA and HST staff involved in the data collection and post-processing/quality assurance undertaken in advance of our analysis.
JBL is supported by an STFC postgraduate studentship.
GMK is supported by the Royal Society as a Royal Society University Research Fellow.
SM is supported by a Research Fellowship from Jesus College, Cambridge.
AB is grateful to the Royal Society for a Dorothy Hodgkin Fellowship. 
PK thanks support from GO-13362 provided by NASA through a grant from STScI under NASA contract NAS5-26555.

\section*{Data Availability}
This work makes use of the following ALMA data: $ADS/$ $JAO.ALMA$ $\#2019.1.01443.T$. 
ALMA is a partnership of ESO (representing its member states), NSF (USA) and NINS (Japan), together with NRC (Canada), MOST and ASIAA (Taiwan), and KASI (Republic of Korea), in cooperation with the Republic of Chile. The Joint ALMA Observatory is operated by ESO, AUI/NRAO and NAOJ.
The HST data used for this study are part of the Cycle 21 with proposal ID GO-13362 and are publicly available on the Barbara A. Mikulski Archive
for Space Telescopes (https://archive.stsci.edu/hst/search.php).
This research has made use of the SIMBAD database, operated at CDS, Strasbourg, France. This research has made use of NASA’s Astrophysics Data System. 
All RV data used to model the planet fits has been included as a supporting file alongside this manuscript.


\bibliographystyle{mnras}
\bibliography{example} % if your bibtex file is called example.bib

%%%%%%%%%%%%%%%%%%%%%%%%%%%%%%%%%%%%%%%%%%%%%%%%%%
%%%%%%%%%%%%%%%%% APPENDICES %%%%%%%%%%%%%%%%%%%%%

\appendix
\section{Fitted Planet Parameters}
\label{sec:AppendixA}
\begin{figure}
    \includegraphics[width=1.0\columnwidth]{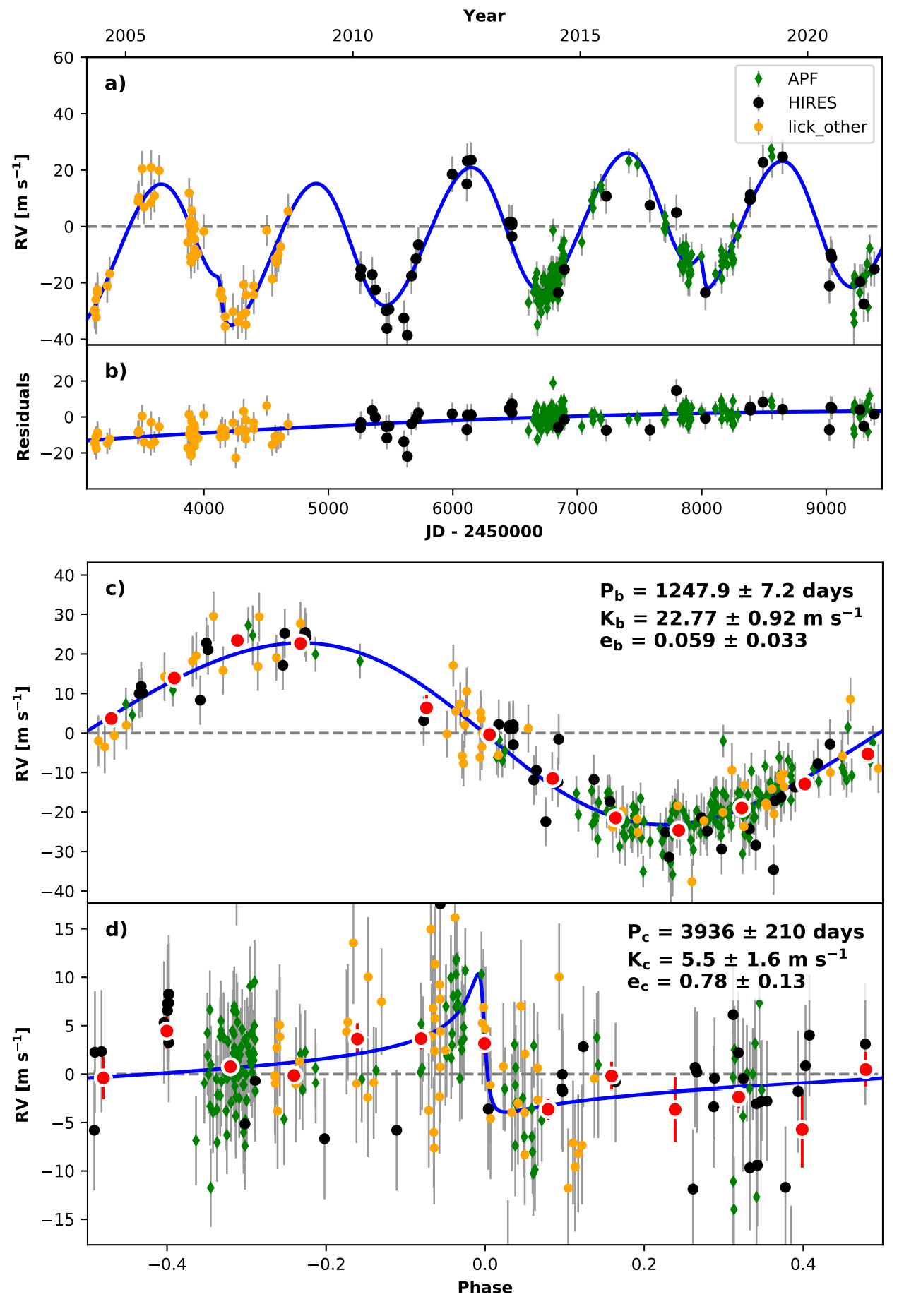}
    \caption{RadVel predictions of the N=2 planet model, based on all available RV data over 17 years.}
    \label{fig:appRadVel}
\end{figure}

Here we provide the plots of best-fit N=2 model derived by RadVel in Fig.~\ref{fig:appRadVel}.

\section{Proper Motion Anomalies: The Long Period Limit}
\label{sec:AppendixB}
In this appendix we develop a simple analytic approximation for the expected proper motion anomaly for a long period binary on a circular orbit. We relate this to observations of the same system in the \textit{Gaia} DR2 and eDR3 catalogues (same start date, with a baseline of 22 and 34 months respectively).

We draw two results from \citet{Penoyre21}, equations 18 and 19. 
Firstly, equation 18 describes the angular separation between the centre of mass and light of an unresolved binary,
\begin{equation}
\bm{\epsilon} = \frac{\varpi a \Delta}{\sqrt{1-\cz_{\phi_v}^2\sz_{i}^2}}\frac{1-e^2}{1+e\cz_\phi}\begin{pmatrix}
\cz_\phi - \cz_{\phi_v-\phi}\cz_{\phi_v}\sz_{i}^2 \\
\sz_{\phi}\cz_{i}
\end{pmatrix}
\label{epsilon}
\end{equation}
where $\varpi$ is the parallax, $a$ the semi-major axis (in au), $e$ is the eccentricity and c and s represent the cosine and sine functions respectively.
The factor $\Delta$ is the relative distance between the center of light and the centre of mass and can be expressed in terms of the mass ratio, $q$, and the light (or flux) ratio, $l$, as
\begin{equation}
\Delta=\frac{q-l}{(1+q)(1+l)}.
\label{delta}
\end{equation}
$\phi$ is the phase of the orbit and $i$ and $\phi_v$ are the inclination and azimuthal viewing angle (such that orbit is closest aligned with the line of sight when $\phi=\phi_v$) respectively.
Secondly, equation 19 from \citet{Penoyre21} describes the average proper motion contribution caused by the orbit of an unresolved binary,
\begin{equation}
\langle \bm{\dot{\epsilon}} \rangle = \frac{1}{t_b-t_a}(\bm{\epsilon}_b-\bm{\epsilon}_a)
\end{equation}
where $t_a$ and $t_b$ are the times of the first and last observation and $\bm{\epsilon}_a$ and $\bm{\epsilon}_b$ are the values of equation \ref{epsilon} at these times. This is a good approximation for the shift in proper motion caused by an unresolved binary and is relatively simple to calculate. 
In the case of a circular orbit equation \ref{epsilon} reduces to
\begin{equation}
\bm{\epsilon} = \frac{\varpi a \Delta}{\Omega}\begin{pmatrix}
\Omega\cz_\phi - \kappa\sz_\phi \\
\sz_{\phi}\cz_{i}
\end{pmatrix}
\label{epsiloncirc}
\end{equation}
where
\begin{equation}
\Omega=\sqrt{1-\cz_{\phi_v}^2\sz_{i}^2}
\end{equation}
and
\begin{equation}
\kappa=\sz_{\phi_v}\cz_{\phi_v}\sz_{i}^2
\end{equation}
are geometric factors entirely dependant on the viewing angle which we can pull out for the moment. The phase of the orbit also obeys a much simpler relationship:
\begin{equation}
\phi=\phi_0 +\frac{2\pi t}{P}
\end{equation}
where $P$ is the period of the orbit.

\subsection{Long period assumption}
If we observe a system from a time $T$ over a time baseline $B$ which is significantly shorter than the period of the binary ($B\ll P$) then the approximate proper motion caused by the binary can be expressed more simply. 
Let $\Phi = \phi(T)$ and $\delta = \frac{2 \pi B}{P} \ll 1$ then
\begin{equation}
\sz_\phi=\sz_\Phi +\delta \cz_\Phi -\frac{\delta^2}{2}\sz_\Phi + O(3)
\end{equation}
and
\begin{equation}
\cz_\phi=\cz_\Phi - \delta \sz_\Phi -\frac{\delta^2}{2}\cz_\Phi + O(3)
\end{equation}
where $O(3)$ denotes terms containing third (or higher) powers of small quantities.
This gives a predicted proper motion contribution of
\begin{equation}
\langle \bm{\dot{\epsilon}} \rangle(\Phi,\delta) = \frac{\varpi a \Delta}{\Omega B}\left[\delta 
\begin{pmatrix} -\Omega^2 \sz_{\Phi} - \kappa \cz_{\Phi} \\ \cz_{i} \cz_{\Phi}\end{pmatrix}
+\frac{\delta^2}{2}
\begin{pmatrix} -\Omega^2 \cz_{\Phi} + \kappa \sz_{\Phi} \\ -\cz_{i} \sz_{\Phi}\end{pmatrix}
\right]+O(3)
\end{equation}

\subsection{Proper motion anomaly}
A single measure of the proper motion of a star cannot differentiate the binary contribution from the linear motion of the system. However if we measure the same system over two or more periods the difference in proper motions can be attributed to a binary companion.
We are interested specifically in comparing proper motions measured in the second and third data releases of the \textit{Gaia} survey. These have the same starting point (i.e. $T$ and $\Phi$ are unchanged) but the baselines differ by a year. We will express this as $B_3=B_2+I$ and $\delta_3=\delta_2+\frac{2\pi I}{P}$ where $I=1$ year (expressed algebraically so that we can see the dimensions of any quantity explicitly).
The proper motion anomaly is
\begin{equation}
\begin{aligned}
\Delta \bm{\mu}=&\langle \bm{\dot{\epsilon}} \rangle(\Phi,\delta_3)-\langle \bm{\dot{\epsilon}} \rangle(\Phi,\delta_2)\\
=&2 \pi^2 \frac{\varpi a \Delta}{\Omega}\frac{I}{P^2}\begin{pmatrix} -\Omega^2 \cz_{\Phi} + \kappa \sz_{\Phi} \\ -\cz_{i} \sz_{\Phi}\end{pmatrix} + O(2).
\end{aligned}
\end{equation}

To translate this into an observable quantity we need to know the orientation of the binary in the reference coordinate system. However even without this information we can calculate the magnitude of the proper motion anomaly
\begin{equation}
|\Delta \bm{\mu}|=2 \pi^2 \frac{\varpi a \Delta I}{P^2}\gamma(\Phi,i,\phi_v)
\end{equation}
where
\begin{equation}
\gamma(\Phi,i,\phi_v)=\frac{1}{\Omega}\left(1-\sz_{i}^2(\cz_{\phi_v}^2+\cz_{\phi_v-\Phi}^2) +\sz_{i}^2 \cz_{\phi_v}^2\cz_{\phi_v-\Phi}^2\right)
\end{equation}
contains all dependence on the initial phase and viewing angle. Note that this result would hold for any I ($\ll P$).
For many companions $q$ is small and $l$ is negligible such that equation \ref{delta} is approximately $\Delta \sim q$. We can also use Kepler's third law to substitute out the period of the system and we find
\begin{equation}
|\Delta \bm{\mu}|\sim \frac{\varpi G M_{pl}}{I a^2}\gamma(\Phi,i,\phi_v)
\end{equation}
where $M_{pl}=qM$ is the planet's mass and $M$ the star's mass.

%
%%%%%%%%%%%%%%%%%%%%%%%%%%%%%%%%%%%%%%%%%%%%%%%%%%
% Don't change these lines
\bsp	% typesetting comment
\label{lastpage}
\end{document}